\documentclass{aa}  

\usepackage{graphicx}
\usepackage{txfonts}
\usepackage{xcolor}
\begin{document} 

   \title{Spatially resolved measurements of the solar photospheric oxygen abundance}

   \author{M. Cubas Armas
          \inst{1},\inst{2}
          \and
          A. Asensio Ramos \inst{1},\inst{2}
          \and
          H. Socas-Navarro\inst{1},\inst{2}}

   \institute{Instituto de Astrof\'isica de Canarias (IAC), Avda V\'ia L\'actea S/N,
                38200 La Laguna, Tenerife, Spain\\
              \email{mcubas\_ext@iac.es}
         \and
             Departamento de Astrof\'isica, Universidad de La Laguna, 38205 La Laguna, Tenerife, Spain\\
              }

   \date{Received ; accepted }

 
  \abstract
   {}
   {We report the results of a novel determination of the solar oxygen abundance using spatially resolved observations and inversions.
   We seek to derive the photospheric solar oxygen abundance with a method that is robust against uncertainties in the model atmosphere. }
   {We use observations with spatial resolution obtained at the Vacuum Tower Telescope (VTT) to derive the oxygen abundance at 40 different spatial positions in granules and intergranular lanes. We first obtain a model for each location by inverting the \ion{Fe}{1} lines with the NICOLE inversion
   code. These models are then integrated into a  
   hierarchical Bayesian model that is used to infer the most probable value for 
   the oxygen abundance that is compatible with all the observations. The abundance is
   derived from
   the [O~{\sc i}] forbidden line at 6300 \AA\, taking into consideration all possible nuisance parameters
   that can affect the abundance. }
   {Our results show good agreement in the inferred oxygen abundance for all the pixels analyzed, demonstrating the robustness of the analysis against possible systematic errors in the model. We find a slightly higher oxygen abundance in granules than in intergranular
   lanes when treated separately ($\log(\epsilon_O)=8.83\pm 0.02$ vs $\log(\epsilon_O)=8.76 \pm 0.02$), which is a difference of approximately 2-$\sigma$. This tension suggests that some systematic errors in the model or the radiative transfer still exist but are small. When taking all pixels together, we obtain an oxygen abundance of 
   log($\epsilon_O$)=8.80 $\pm$ 0.03, which is compatible with both granules and lanes within 1-$\sigma$. The spread of results is due to both systematic and random errors. }
   {}

   \keywords{Sun: abundances --
                Sun: atmosphere --
                Sun: photosphere --
                methods: statistical
               }

   \maketitle
%
\section{Introduction}

In the mid-2000s, Asplund and collaborators published a series of papers \citep[e.g.,][]{2001ApJ...556L..63A,2004A&A...417..751A,2006A&A...456..675S,2009ARA&A..47..481A}
claiming that the solar chemical composition should be revised to a lower metallicity ($Z$). 
The new lower abundances present some advantages, such as a better fit with the interstellar 
medium; however, this has been questioned by \citet{2006A&A...445..633E}, who claim that the composition of planet-hosting stars should be considered. On the other hand, these new abundances ruin the existing agreement between the predictions of solar interior models and helioseismic measurements \citep[see][]{2008PhR...457..217B}. 

Besides the studies of Asplund and collaborators, there were other groups that worked on the solar abundances, such as Caffau and collaborators, who published another series of papers using the CO\textsuperscript{5}BOLD simulations \citep{2012JCoPh.231..919F}. These latter authors obtained much larger abundances \citep[e.g.,][]{2008A&A...488.1031C, 2010A&A...514A..92C,2011SoPh..268..255C, 2013A&A...554A.126C}, showing that 3D hydrodynamical simulations do not necessarily suggest low metallicity. There is still no clear consensus on what set of abundances is more realistic and the issue of the solar chemical composition 
remains largely unresolved. 

In the meantime, modelers have taken a critical look at 
solar interior theories to explore possible modifications capable of 
reconciling the models with low-Z abundances. In doing so, the opacity 
data used to construct the models were found to be largely outdated. More recent calculations yield 
significantly higher opacities, which helps to absorb the impact of a lower metallicity \citep[see][]{2009A&A...494..205C,bailey}. 
The issue with opacities has become even more troublesome with new laboratory 
measurements of iron under a particular set of stellar interior conditions, 
resulting in values significantly higher than those of the most sophisticated 
tabulations \citep{bailey}. However, even if one could somehow tune the opacities 
to come into agreement with a low-Z solar composition, there would still be unresolved
conflicts with incompatible solar wind measurements and also with results from 
neutrino experiments \citep[e.g.,][]{2016ApJ...816...13V,2008ApJ...687..678H}.

In order to resolve the current uncertainties discussed above, more robust empirical constraints on the solar chemical composition are needed. In this paper we focus on the case of oxygen abundance, which is both extremely difficult to constrain and crucially important. This difficulty stems from the fact that there are very few spectral lines in the solar spectrum 
and they are all either weak, blended with other lines, or affected by complex 
nonlocal thermodynamic equilibrium (NLTE) formation. 
Oxygen abundance is crucial because: (i) it is the most abundant of all metals in the Universe; (ii) it 
is one of the main contributors of free electrons and interior opacity; and (iii) some 
of the other important elements for interior models (especially  N, Ne and Ar) cannot 
be directly measured (or only with much increased difficulty) and are typically measured 
relative to O. Therefore, the so-called {solar oxygen crisis} \citep{2012AAS...21914408A} 
forms a fundamental part of the chemical composition problem. One of the main difficulties is 
that the abundances determined from the interpretation of stellar spectra require 
the prescription of a model atmosphere. The older high-Z abundances were obtained with
one-dimensional (1D) semi-empirical  models \citep[i.e.,][]{1998SSRv...85..161G}. The Asplund 
et al. low-Z abundances are obtained from a three-dimensional (3D) hydrodynamical simulation of 
the atmosphere. \cite{2008ApJ...686..731A} argued that, while 3D is clearly a better 
choice than 1D, it is not clear that, when one is fitting synthetic spectra to 
observations, a theoretical ab initio model should be preferred over a semi-empirical 
model. This is consistent with the observations because the 1D shortcomings would have been absorbed by the fitting parameters. \cite{2015A&A...577A..25S} proposed 
to overcome the dilemma between 1D empirical and 3D theoretical models by using 3D 
empirical models. In the same paper, he assessed the importance of systematic effects, concluding 
that they could be responsible of log($\epsilon_O$) fluctuations between 8.70 and 8.90 
(i.e., from intermediate to high-Z values). Seemingly minor details, such 
as minuscule variations in the placement of the continuum or slight changes in the 
absolute wavelength calibration, have a very large impact on the final result. It is therefore very difficult to make conclusive statements on the oxygen abundance. Independent evidence is needed to make progress, ideally from alternative strategies that are less sensitive to the practical details of the data processing or the (prescribed) model atmosphere.

Here we present a new approach to the problem using spatially resolved observations. 
We carried out slit-spectroscopy observations in Fe~{\sc i} and 
[O~{\sc i}] spectral lines simultaneously, thus collecting spectral profiles at many different 
spatial locations, including granules and intergranular lanes, spanning a temperature 
range of nearly 1000K in the photosphere. The approach employed here draws the best 
of both approaches previously discussed: (i) the model is empirical, obtained by fitting 
the observed Fe~{\sc i} lines, and (ii) the model is 3D because each spatial pixel 
has its own temperature, density, and velocity stratification. Another novel ingredient 
in this approach with respect to previous works is that we do not compare with the 
average solar spectrum from the FTS atlas observation to derive the overall solar oxygen
abundance. Instead, we fit each individual profile observed at each pixel using the same
instrument and configuration as that used to derive the atmosphere, therefore obtaining  
an oxygen abundance value for each point.

\section{Observations}

Our observations were acquired with the Vacuum Tower Telescope
\citep[VTT][]{1985VA.....28..519S,1998NewAR..42..493V} at the Observatorio de 
Iza\~na, Tenerife, Spain. The VTT is a solar telescope with two coelostat mirrors at the 
top of a building of approximately 38 m in height. The coelostat 
mirrors deliver a nonrotating image of the Sun. The telescope primary 
mirror has a diameter of 70 cm and a focal length of 46 m. The VTT is equiped with 
an adaptive-optics system
\citep[KAOS,][]{2002AN....323..236S,2003SPIE.4853..187V} with a wavefront sensor, a 
deformable mirror, and a high-speed camera. 

We use the Echelle spectrograph, one of the available optical instruments at the
VTT. It consists of a grating spectrograph with a resolving power of $R \approx 10^6$. The slit covers about 200$"$, 
equivalent to roughly 145 Mm on the Sun. Moreover, the VTT is equipped with 
a slit-jaw camera for observations in white light, Ca K, and H$\alpha$. For more information on the telescope, see \cite{VONDERLUHE1998493}.

\begin{figure} 
\centering
\includegraphics[trim={0cm 1.cm 0cm 2.cm},clip,width=1\hsize]{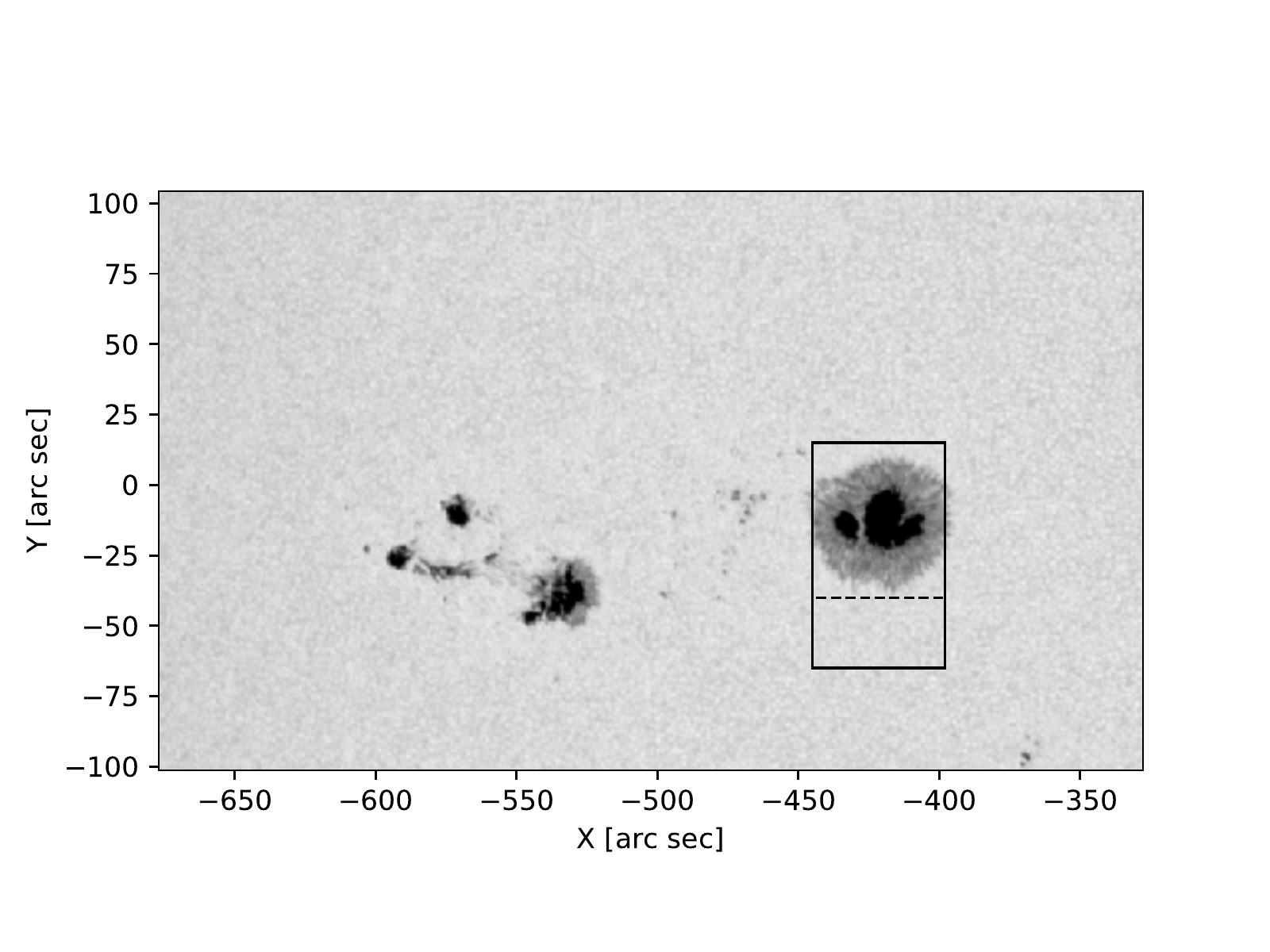} 
\caption{Continuum context image of active region AR12565 on July 15 2016. Image from the HMI instrument onboard SDO \citep{2012SoPh..275..207S}. Our observations are represented by the solid rectangle in the image. The area under the dashed line corresponds to the data employed in this work.}
\label{fig:map}
\end{figure}

The observational data used in this work were acquired on July 15 2016 starting at 
11:34 UTC, with an exposure time per slit position of 200 ms. The scanning time to 
cover the entire map was around 4 min (plus the overhead time
for data saving and slit motion). The observations produce a 3D $(x,y,\lambda)$
cube with the spectrum for each pixel in the 2D field of view (FOV) in a quiet region 
adjacent to active region AR12565 (see the rectangle plotted in Fig.~\ref{fig:map}), located in the solar equator at around 20º west of the solar disk center. Pores in the nearby active region were used as targets for the 
KAOS adaptive optics system to lock onto. Following standard procedures, we also took dark-current and flat-field images. The standard data-reduction procedure was applied, consisting 
of dark-current subtraction and flat-field normalization to correct for 
pixel-to-pixel sensitivity variations. The flat-field images are produced, as usual in this facility, by acquiring
and summing images while the telescope is moving on the quiet Sun. This produces an image with no spatial structure from which a good flat-field is obtained after removing the spectrum. Wavelength calibration and 
correction of the prefilter curvature were also applied. Absolute wavelengths were 
determined by fitting an average disk-center quiet-Sun profile to the FTS atlas
\citep{1999SoPh..184..421N}. The same fit provides the calibration between counts and physical
units and an estimate of the amount of stray light in the observations. Continuum 
rectification is done by fitting a third-order polynomial to a few selected continuum 
windows. We estimate the noise in the observations to be around 10$^{-3}$ in units of the continuum intensity.

\begin{figure*} 
\centering
\includegraphics[trim={3.7cm 0cm 3.5cm 1.2cm},clip,width=0.4\textwidth]{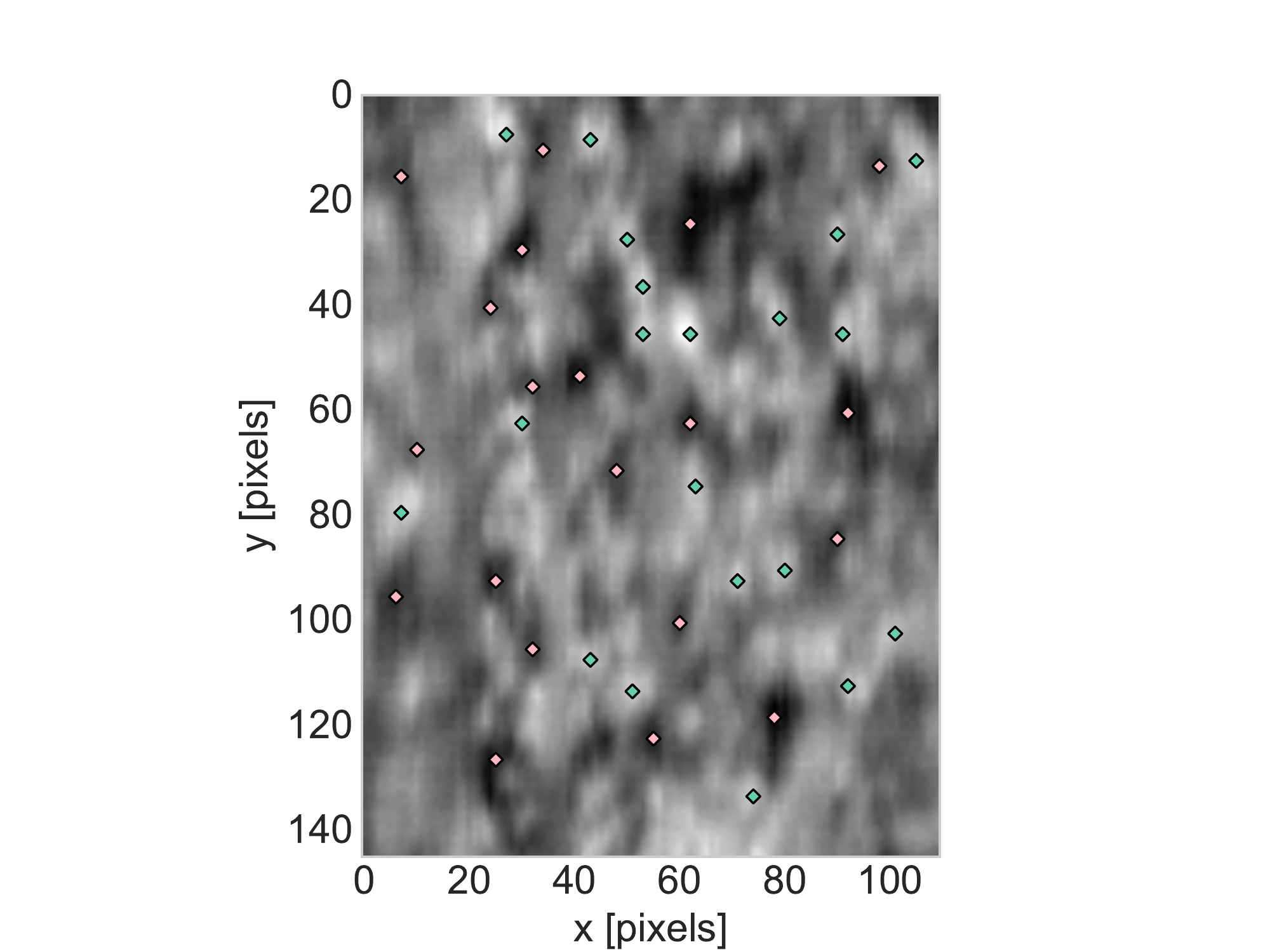}
\includegraphics[trim={0.2cm 0.0cm 0cm 0.cm},clip,width=0.5\textwidth]{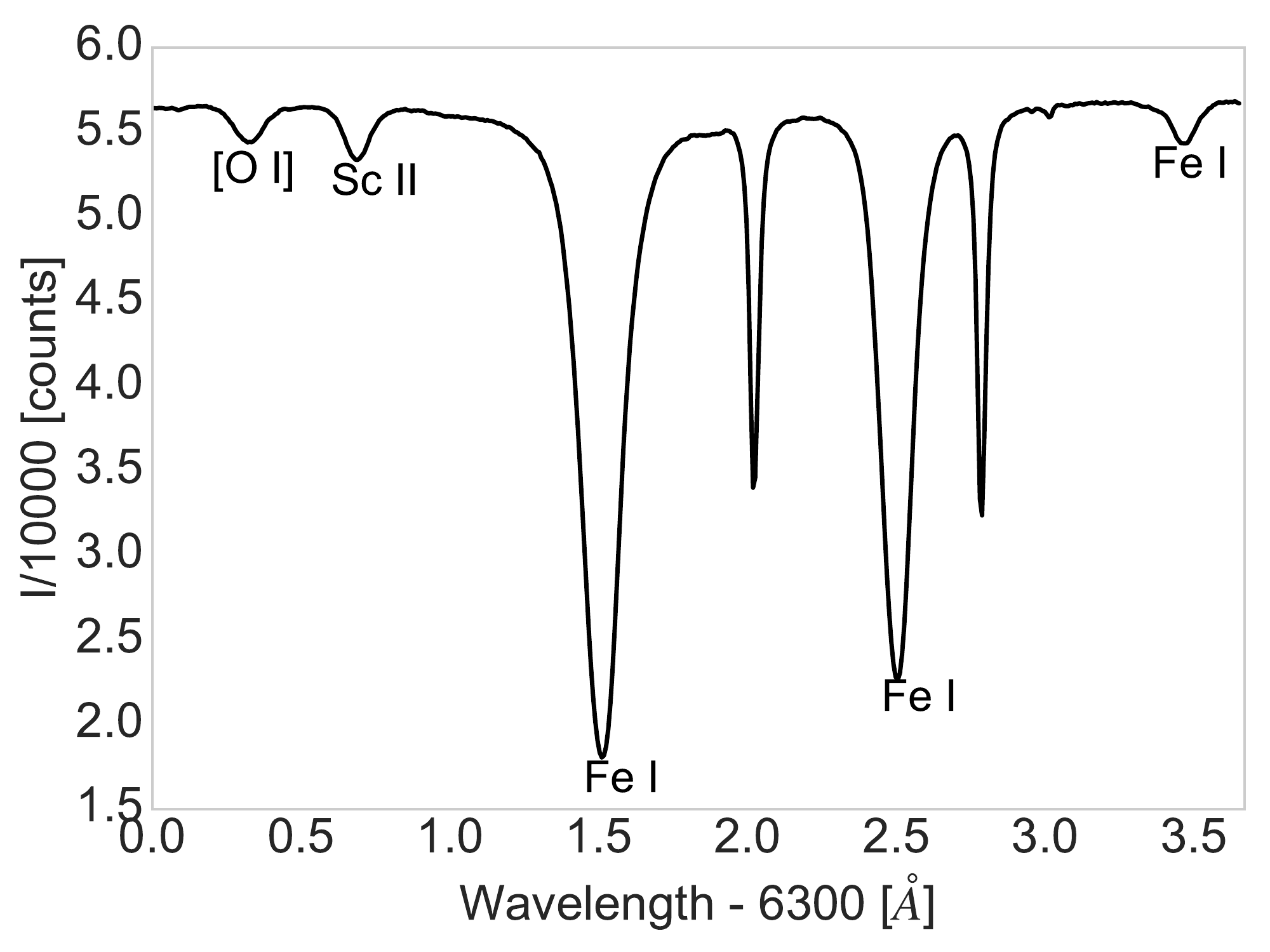}
\caption{Left: Continuum map, marking the position of the pixels considered for this study: granules in green and intergranular lanes in pink (40 pixels in total). As standard practice, these pixels were
selected for their continuum brightness, choosing the brightest and darkest points in the FOV. Right: the average spectrum in the FOV shows the lines of interest in our work.}
\label{fig:profile}
\end{figure*}

A 2D map at a continuum wavelength in the 6300 \AA\ region and the mean spectrum in 
the region are plotted in Fig.~\ref{fig:profile}. The spectral lines of interest are, from 
left to right: the [O~{\sc i}] forbidden line that we use to infer the oxygen abundance,
a Sc~{\sc ii} line used for the calibration of the missing dynamics in the inversions 
(see following section), and the three Fe~{\sc i} lines that we used to construct the model 
atmospheres, the two widely used Fe~{\sc i} lines at 6301.5 \AA{} and 6302.5 \AA{}, 
and a weak Fe~{\sc i} line at 6303.5 \AA{}. The two narrow spectral lines at 
6302.0 \AA{} and 6302.7 \AA{} are telluric contamination from the Earth's atmosphere.

\section{Model atmospheres}

We carry out inversions of the observed spectra using the NICOLE inversion code
\citep{2015A&A...577A...7S}. 
The inversions are done independently for all 40 spatial locations, corresponding to those pixels marked 
in Fig.~\ref{fig:profile}. Granules are marked in green while intergranular lanes are marked in pink. We did not use all pixels of  
the image to keep computational time within reasonable values but this sample provides results
with statistical relevance. The pixels were selected for their brightness in the continuum, choosing them 
from among the brightest (for granules) and darkest (for lanes) in the FOV. This extreme selection 
will enhance any possible difference between granules and lanes. With this 
strategy, we aim to obtain 40 different determinations of the oxygen abundance which can then be 
compared to one another. As the inversion is done pixel by pixel, the resulting model is 
assumed to be spatially resolved with a spatial resolution of 0.8$"$/px. The pair of lines 
at 6301.5 and 6302.5 \AA\ are relatively strong lines. As already pointed out by
\cite{2015A&A...577A..25S}, this induces a reduced sensitivity to the velocity in the lower 
layers of the atmosphere, near the surface $\tau_\mathrm{cont}=1$ for continuum wavelengths, precisely where
the [O~{\sc i}] line is formed. For that reason, we add the weak Fe~{\sc i} line at 
6303.5 \AA\ to constrain the relevant dynamics that might have been missed by the stronger lines. This is an
advantage conferred by our observations with respect to those of Hinode used by 
\cite{2015A&A...577A..25S}, which did not reveal the weak Fe \textsc{i} line. 
The atomic data used for all transitions in the spectral range are summarized in Table~\ref{table:lines2}.

\begin{table*}
\centering                          
\begin{tabular}{c c c c c c c c c}        
\hline\hline                 
Ion &  $\lambda$ [\AA{}] & $\chi_e$ [eV] & log(gf) & $\gamma_{rad}$  & $\gamma_{Stark}$ & $\gamma_{Waals}$ & $\sigma$ & $\alpha$\\
\hline  
[O~{\sc i}]  & 6300.304  & 0.000 & -9.717$^a$ & 0.0 & 0.05  & 1.00 & ... & ...\\
$^{58}$Ni~{\sc i}   & 6300.335  & 4.266 & -2.253$^b$ & 2.63& 0.054& 1.82 & ... & ...\\
$^{60}$Ni~{\sc i}   & 6300.355  & 4.266 & -2.663$^b$ & 2.63& 0.054& 1.82 & ... & ...\\
Sc~{\sc ii} & 6300.678   & 1.507 & -1.970 & 2.30& 0.05 & 1.30 & ... & ...\\
Fe~{\sc i}  & 6301.501 & 3.654 & -0.718$^c$ & ... & ... & ... & 834.4 & 0.243\\
Fe~{\sc i}  & 6302.494 & 3.686 & -1.25 & ... & ... & ... & 850.2 & 0.239\\
Fe~{\sc i}  & 6303.462 & 4.320 & -2.66  & 2.40 & 0.11 & 1.95 & ... & ...\\
 \hline                                   
\end{tabular}
\caption{Adopted atomic parameters. $\chi$ is the excitation potential. $\gamma_{rad}$, $\gamma_{Stark}$ and $\gamma_{Waals}$ are the radiative, the Stark, and the van der Waals damping parameters (units $10^8$ $rad$ $s^{-1}$). The log($gf$) are taken from: $^a$ \cite{2000MNRAS.312..813S}, $^b$ \cite{2003ApJ...584L.107J}, $^c$ VALD database. In the case of the Fe~{\sc i} lines, the $\gamma_{rad}$ is introduced by NICOLE as calculated in \cite{1976oasp.book.....G}. The log(gf) values for the Sc~{\sc ii} and Fe~{\sc i} 6302.5 \AA{} lines have been revised in this paper.}   
\label{table:lines2}      
\end{table*}

\begin{figure}
\centering
\includegraphics[trim={0cm 0cm 1cm 1cm},clip,width=.99\hsize]{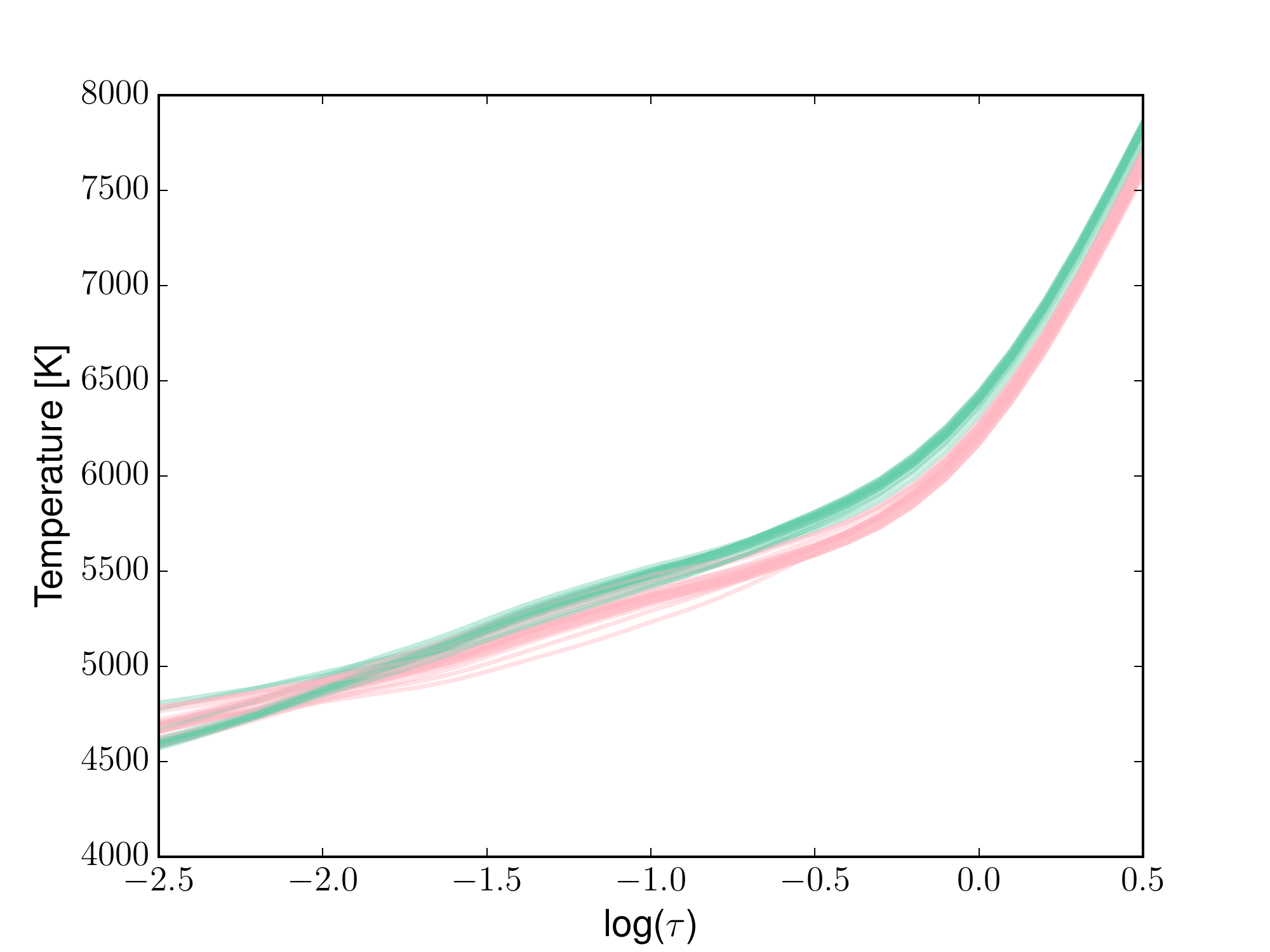}
\caption{Temperature stratification of each inferred model. The optical depth scale is measured at 5000 \AA{}.}
\label{fig:T}
\end{figure}

\begin{figure}
\centering
\includegraphics[trim={0.5cm 0cm 0cm 0cm},clip,width=1\hsize]{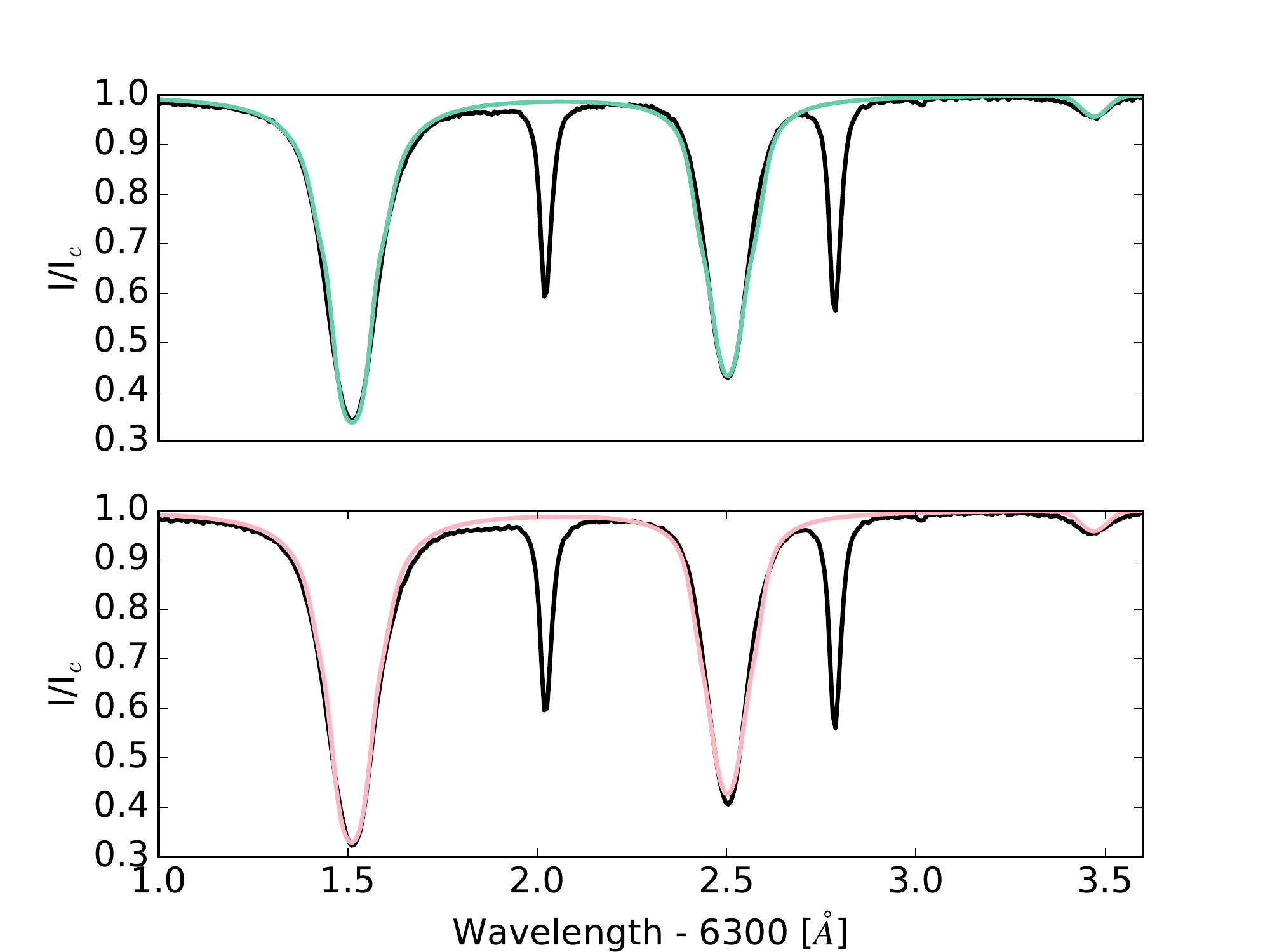} 
\caption{Inversions of two spatial pixels. Upper panel: Spectral profile of a granule (black) and fit (green). Lower panel: Spectral profile of an intergranular lane (black) and fit (pink). I$_c$ is the average quiet Sun continuum intensity. }
\label{fig:invers}
\end{figure}

The inversions were performed using eight nodes for the temperature and one for the Doppler 
velocity. We did not invert the magnetic field because our observations do not include polarimetry. As we deal with quiet Sun profiles, we can assume that magnetic fields 
do not appreciably affect the intensity profiles. This turns out to be a good approximation according to
\cite{2008ApJ...673..470B} and \cite{Fabbian+Moreno-Insertis2015}. We use six nodes for 
the depth stratification of the microturbulence. This turns out to be important because the 
synthetic Sc~{\sc ii} 
line close to the [O\textsc{i}] feature in inverted models appears 
too deep when compared with solar atlases. This points to a lack of small-scale dynamics at the base 
of the photosphere, where these lines are 
formed. Consequently, adding a depth stratification of the microturbulence in the model accounts for 
the missing broadening in the deepest layers. The inferred temperature stratifications of all considered 
models are displayed in Fig.~\ref{fig:T}. Granular models are shown in green, while those of 
lanes are displayed in pink. All models share similar temperature stratifications 
in the range of $\log \tau_{5000}$ (with $\tau_{5000}$ the optical depth measured at 5000 \AA{}) between 
-2.5 and 0.5, which is where the iron lines that are used here are formed. However, we find that
granules display slightly higher temperatures than lanes through the O and Ni line-formation region, although a cross-over point is found in the high photosphere (this gives rise to the reverse-granulation effect seen in the cores of strong lines). We note here that our model extends from $\log \tau_{5000}$=-7 to $\log \tau_{5000}$=2, but information that can be used to constrain regions below $\tau_{5000}>0.5$ and $\tau_{5000}<-2.5$  is scarce in the spectrum.
Examples to illustrate the good fit quality are given in Fig.~\ref{fig:invers}.

Once we have inferred model atmospheres for all 40 pixels, we compute a grid of synthetic spectra by keeping them fixed while modifying the solar oxygen abundance in the 
range $[8.50, 9.20]$ and the nickel abundance in the range $[5.80, 6.36]$. Considering the
Ni \textsc{i} line is crucial because it is blended with the [O~{\sc i}] line. We included 
the two major isotopes $^{58}$Ni~{\sc i} and $^{60}$Ni~{\sc i}   in the calculation as in \cite{2003ApJ...584L.107J}. We uniformly 
sample both intervals of abundances  with a step of 0.05 for oxygen 
and 0.04 for nickel. The ranges include high and low abundance values as reported 
in the literature \citep[e.g.,][]{1989GeCoA..53..197A,1998SSRv...85..161G,2009ARA&A..47..481A}. We note that this work uses NICOLE (both in inversion and synthesis mode) under LTE conditions. The results 
of the synthesis are used as a precomputed database to carry out Bayesian inference using a simulator
\citep{ohagan06}. This simulator uses a simple bilinear interpolation to synthesize the spectra
for arbitrary values of the oxygen and nickel abundances.

\section{Bayesian analysis}
 In the following we describe the Bayesian models \citep[e.g.,][]{gregory05} that we use to 
obtain information about the oxygen abundance. To this end, we compute the marginal posterior probability distribution of the oxygen abundance in 
all 40 pixels. The first ingredient that we need to define is the generative model, from which
the likelihood function will naturally emerge. We model the observed spectrum by a synthetic
spectrum obtained under the LTE approximation in 
the atmospheric models obtained from the inversion. We augment the parameters of the models
with some nuisance parameters:

\begin{equation}
I_{\mathrm{obs,i}}(\lambda) = c_i I_{\mathrm{syn,i}} \left( \lambda - \lambda_0 \frac{w_i}{c}; \mathbf{p}_i \right)+\epsilon_i,
\end{equation}
where $i=1,\ldots,N$ represents the label for each one of the $N$ pixels and $\lambda_0$ is the central
wavelength of the spectral line of interest. The synthesis depends on the set
of model parameters $\mathbf{p}_i=\{T_i,v_i,\log(\epsilon_O)_i,\log(\epsilon_{Ni})_i\}$. Here, $T$ and
$v$ are the temperature and velocity obtained from the inversions, respectively. We also 
find the oxygen abundance, that
we denote $\log(\epsilon_O)$, and the nickel abundance,
represented by $\log(\epsilon_{Ni})$. Furthermore, we allow for an additional
wavelength shift of the line due to inaccuracies in the wavelength scale, which we
encode in a Doppler velocity $w$. Finally, we rescale the continuum by a 
certain correction factor $c,$ which is used to set all profiles to the same continuum level.
The synthetic spectrum is perturbed with noise $\epsilon,$ which we assume to be Gaussian
with diagonal covariance matrix and variance $s^2$.
This assumption automatically defines the likelihood for each pixel, which is given by
the product of $N_\lambda$ uncorrelated normal distributions, with $N_\lambda$ being
the number of wavelength points considered in the spectrum:

\begin{align}
p(D|\mathbf{p}_i,&c_i,s_i) = \frac{1}{s^{N_\lambda} (2\pi)^{{N_\lambda}/2}}\\
& \prod_{j=1}^{N_\lambda} \exp \left\{-\frac{1}{2s^2} \left[ I_{\mathrm{obs,i}}(\lambda_j) - c_i I_{\mathrm{syn,i}} \left( \lambda_j - \lambda_0 \frac{w_i}{c}; \mathbf{p}_i \right) \right] \right\},
\end{align}
where $D$ represents all the observed data.
All parameters except the oxygen abundance are considered as nuisance parameters 
and are marginalized out at the end. For computational reasons, we consider the 
temperature and velocity of the model as fixed quantities and we leave the fully Bayesian 
approach for a future work. Consequently, we take their priors to be Dirac deltas, which have 
no effect when marginalized out. The specific priors selected in this work for the rest of 
parameters are specified in Table~\ref{table:priors1}. Since we have no clear preference for 
any value of the oxygen abundance, we choose a flat prior. On the contrary, the nickel abundance 
is very well determined from previous works and we choose a Gaussian prior with a mean that is 
equal to the consensus value and a small dispersion according to the results 
of \cite{2009ApJ...691L.119S}. The posterior distribution, following the Bayes theorem, is:

\begin{equation}
p(\mathbf{p},\mathbf{c},\mathbf{s}|D)   \propto \nonumber 
p(\mathbf{p},\mathbf{c},\mathbf{s})  p(D|\mathbf{p},\mathbf{c},\mathbf{s}),
\end{equation} 
where $\mathbf{p}=\{\mathbf{p}_1,\ldots,\mathbf{p}_N\}$, $\mathbf{c}=\{c_1,\ldots,c_N\}$ and 
$\mathbf{s}=\{s_1,\ldots,s_N\}$. We assume that the prior factorizes for all
pixels, meaning that

\begin{equation}
p(\mathbf{p},\mathbf{c},\mathbf{s})= \prod_{i=1}^N p(\mathbf{p}_i)  p(c_i)  p(s_i).
\end{equation}

\begin{table}
\centering                          
\begin{tabular}{lll}        
\hline\hline                 
Parameter &  Range & Type  \\
\hline  
T$_i$ & --- & $\delta$(T$_i$ - T$_{model}$)\\
v$_i$ & --- & $\delta$(v$_i$ - v$_{model}$)\\
log($\epsilon_O$)  & [8.50, 9.20] & Uniform \\
log($\epsilon_{Ni}$)  & [5.80, 6.36] & Bounded normal $\mu$=6.17; $\sigma$=0.05$^a$\\ 
w [km/s] & --- & Normal $\mu$=0; $\sigma$=2\\
c & --- & Normal $\mu$=1; $\sigma$=0.2\\
log(s) & [-3,6] & Uniform\\
 \hline                                   
\end{tabular}
\caption{Prior selected for each parameter of our unpooled Bayesian analysis. $^a$taking into account \cite{2009ApJ...691L.119S}.}    
\label{table:priors1}      
\end{table}

We explore two different Bayesian models for our data. In the first one, each pixel is considered 
independent of the rest (also known as unpooled model), obtaining the marginal posteriors distribution 
for the oxygen abundance of each pixel. In the second model, the oxygen abundance is allowed to 
vary from pixel to pixel but all of them are extracted from a common prior, a hierarchical 
partial pooling model. As defined in Table \ref{table:priors2}, this common prior is chosen 
to be Gaussian, with hyperparameters $\mu_O$ and $\sigma_O$. 
These parameters are then interpreted as a global estimation of
the mean oxygen abundance together with its variability in the
pixels that we considered. The advantage of a hierarchical pooling model
lies in the fact that we aggregate all pixels while simultaneously
allowing for pixel-by-pixel variations,
wherever they exist. In this case, the prior 
for $\log(\epsilon_O)$ depends on the hyperparameters $\mu_O$ and $\sigma_O$, meaning that 

\begin{equation}
    \prod_{i=1}^N p(\log(\epsilon_O)_i,\mu_O,\sigma_O) = p(\mu_O)  p(\sigma_O) \prod_{i=1}^N p(\log(\epsilon_O)_i|\mu_O,\sigma_O),
\end{equation}
and $\mu_O$ and $\sigma_O$ are added as parameters during the inference. We note that in this case, the parameter $s$ is also a common prior. 

\begin{figure}
\centering
\includegraphics[trim={0cm 0cm 0cm 0cm},clip,width=.95\hsize]{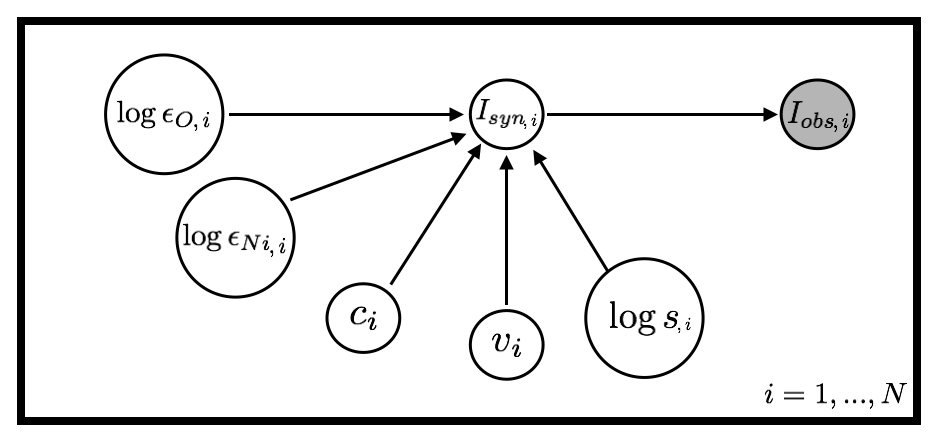} 
\includegraphics[trim={0cm 0cm 0cm 0cm},clip,width=.95\hsize]{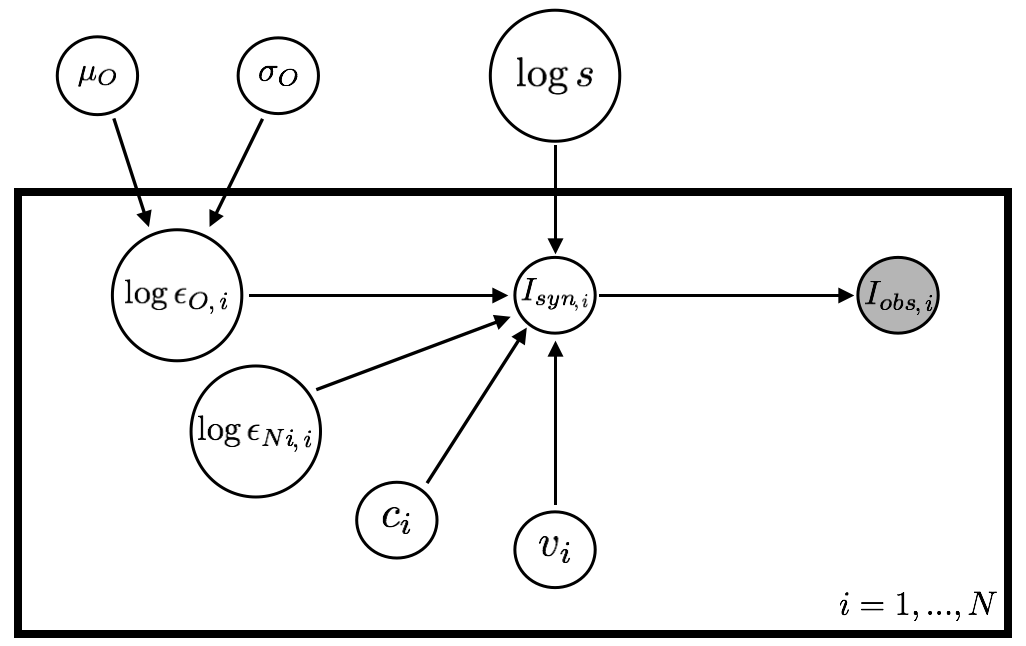} 
\caption{Representation of the parameters for the unpooled model ({\it upper panel}) and the hierarchical partial pooling model ({\it bottom panel}). For each pixel, we have four parameters that contribute to the shape of the synthetic intensity profile. In addition to those parameters, in the hieararchical partial pooling model, we have two hyperparameters, the mean value and the standard deviation of the Gaussian, which represent the global oxygen distribution. The parameter $s$ accounts for the uncertainties of our model.}
\label{fig:esquema}
\end{figure}

The two probabilistic models are displayed in graphical form in Fig.~\ref{fig:esquema} (upper 
panel corresponding to the unpooled model and bottom panel to the hierarchical partial pooling model), 
where all conditional dependences are shown as directed links.

\begin{table}
\centering                          
\begin{tabular}{lll}        
\hline\hline                 
Parameter &  Range & Type\\
\hline 
T$_i$ & --- & $\delta$(T$_i$ - T$_{model}$)\\
v$_i$ & --- & $\delta$(v$_i$ - v$_{model}$)\\
$\mu_O$ & [8.40, 9.10] & Uniform \\
$\sigma_O$ & $\sigma$=1 & Half Normal \\
log($\epsilon_O$)  & [8.40, 9.10] & Bounded normal $\mu$=$\mu_O$; $\sigma$=$\sigma_O$\\
log($\epsilon_{Ni}$)  & [5.80, 6.36] & Bounded normal $\mu$=6.17; $\sigma$=0.05$^a$\\ 
v(ws) [\AA{}] & [-0.2, 0.2] & Bounded normal $\mu$=0; $\sigma$=0.1\\
c factor & --- & Normal $\mu$=1; $\sigma$=0.2\\
log(s) & [-3,6] & Uniform\\
 \hline                                   
\end{tabular}
\caption{Prior selected for each parameter of the Bayesian analysis using a hierarchical partial pooling model.$^a$taking into account \cite{2009ApJ...691L.119S}.}    
\label{table:priors2}      
\end{table}

The sampling of the posterior is done with the PyMC3 Python package \citep{pymc3}, which is designed for 
Bayesian statistical modeling and uses advanced Markov Chain Monte Carlo (MCMC) 
sampling algorithms 
to generate samples from the posterior distribution. Since the sampling of hierarchical 
Bayesian models 
is especially difficult and prone to problems, we relied on the 
no-U-turn (NUTS, \citealp{hoffman})
Hamiltonian Monte Carlo-type (HMC) method.
The NUTS sampler automatically tunes the parameters of the HMC algorithm in 
order to allow for an efficient
sampling of the posterior. It works well on high-dimensional and complex posterior 
distributions\footnote{For a precise definition of the NUTS algorithm, 
see \cite{2012arXiv1206.1901N}, \cite{hoffman} and \cite{2017arXiv170102434B}.}.

\begin{figure*} 
\centering
\includegraphics[trim={0cm 0cm 0cm 0cm},clip,width=0.86\hsize]{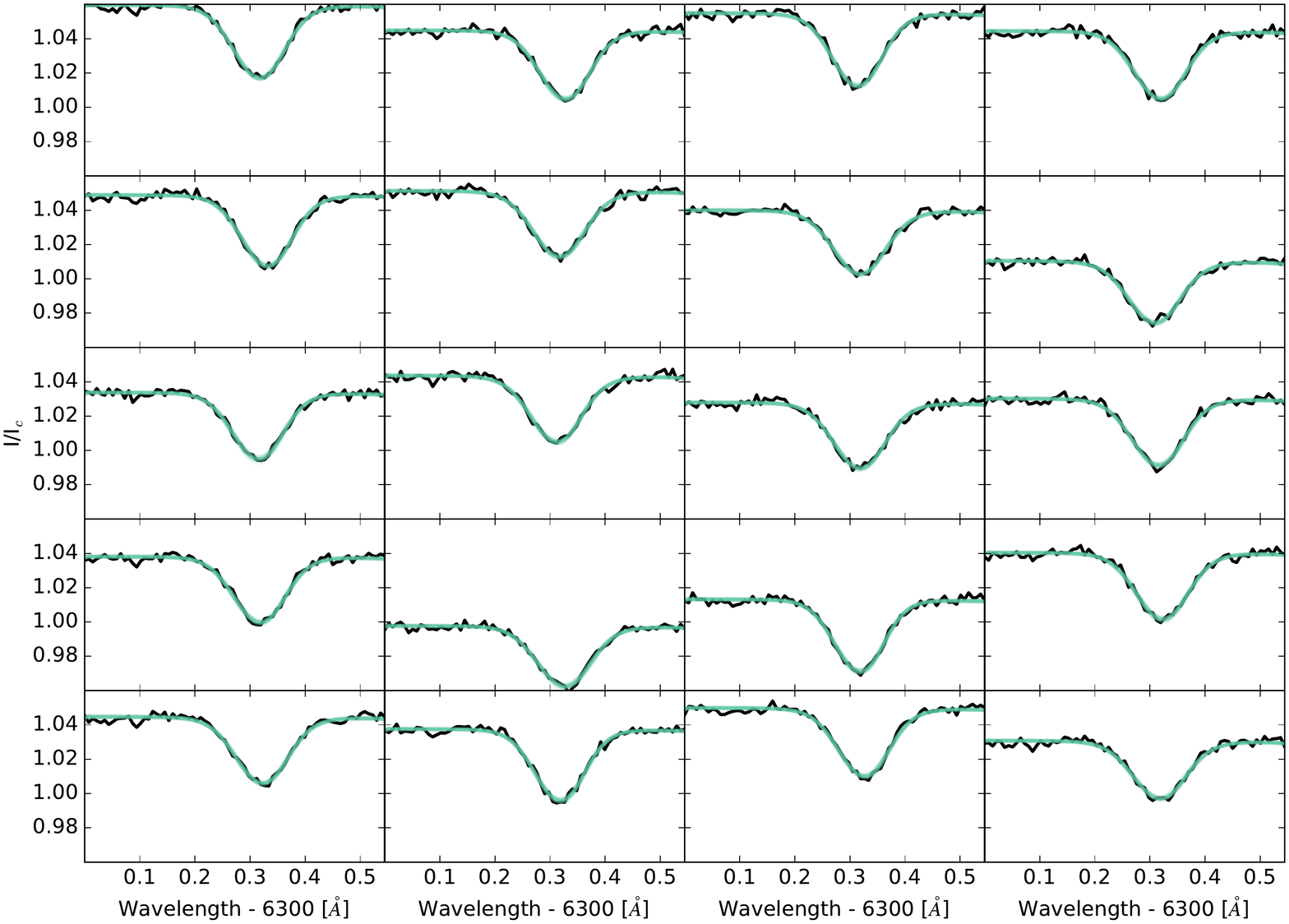} 
\caption{Observed (black) and synthetic (green) profiles using values extracted from the posterior for the granules in the unpooled model. The continuum taken to normalize the intensity is the mean continuum between granules and lanes pixels.}
\label{fig:bay_gran}
\end{figure*}

\begin{figure*} 
\centering
\includegraphics[trim={0cm 0cm 0cm 0cm},clip,width=0.86\hsize]{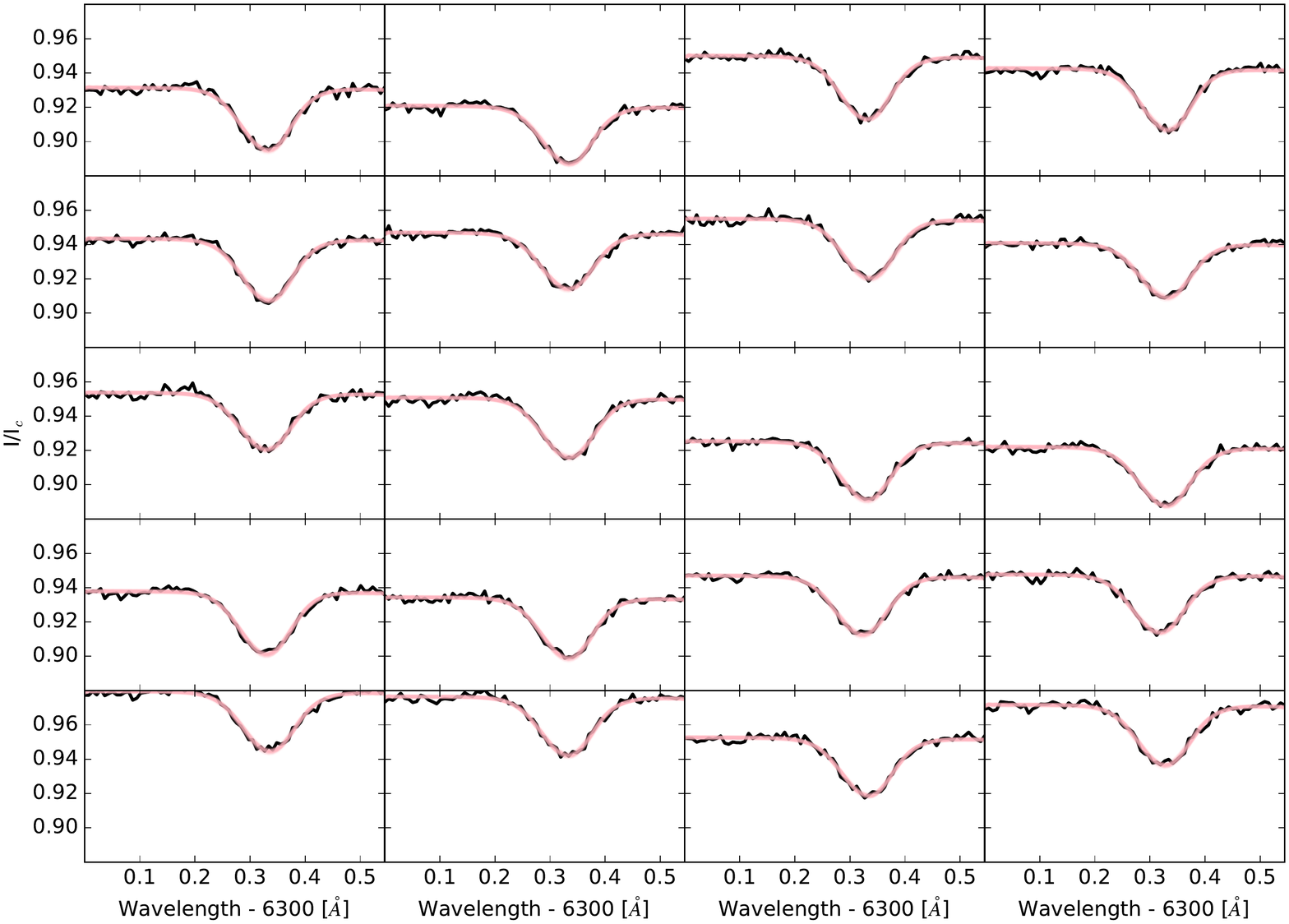} 
\caption{Observed (black) and synthetic (pink) profiles using values extracted from the posterior for the lanes in the unpooled model. The continuum taken to normalize the intensity is the mean continuum between granules and lanes pixels.}
\label{fig:bay_lane}
\end{figure*}

Once samples from the posterior are obtained, it is always good practice to 
produce posterior predictive checks 
in which the posterior is used to produce synthetic [O~{\sc i}] line profiles.
These checks will clearly show whether or not the generative model is a representative model
of the observations.
Figure \ref{fig:bay_gran} displays posterior predictive checks for all pixels in
granules in the unpooled model. Correspondingly, Fig. \ref{fig:bay_lane} shows the same 
for intergranular lanes. The noisy observations are displayed in black. The ability of 
our model to explain the observed [O~{\sc i}] line is remarkable. Similar plots are 
presented in the Appendix for the hierarchical partial pooling model for granules, intergranules,
and both granules and intergranules together.

%
\section{Results and Discussion}

\subsection{Unpooled model}

\begin{figure} 
\centering
\includegraphics[trim={3 10 5 7},clip,width=.95\hsize]{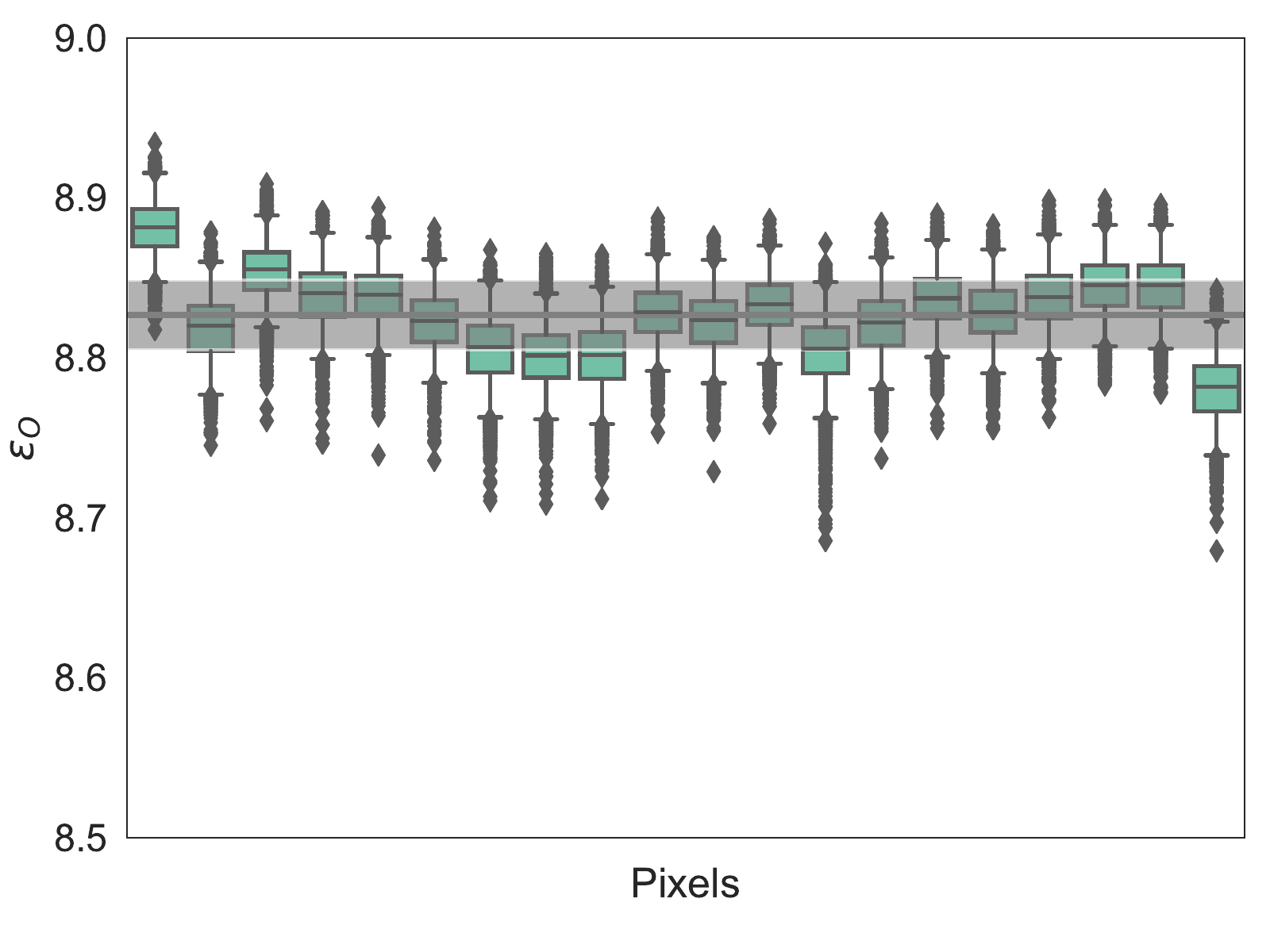}
\includegraphics[trim={3 10 5 7},clip,width=.95\hsize]{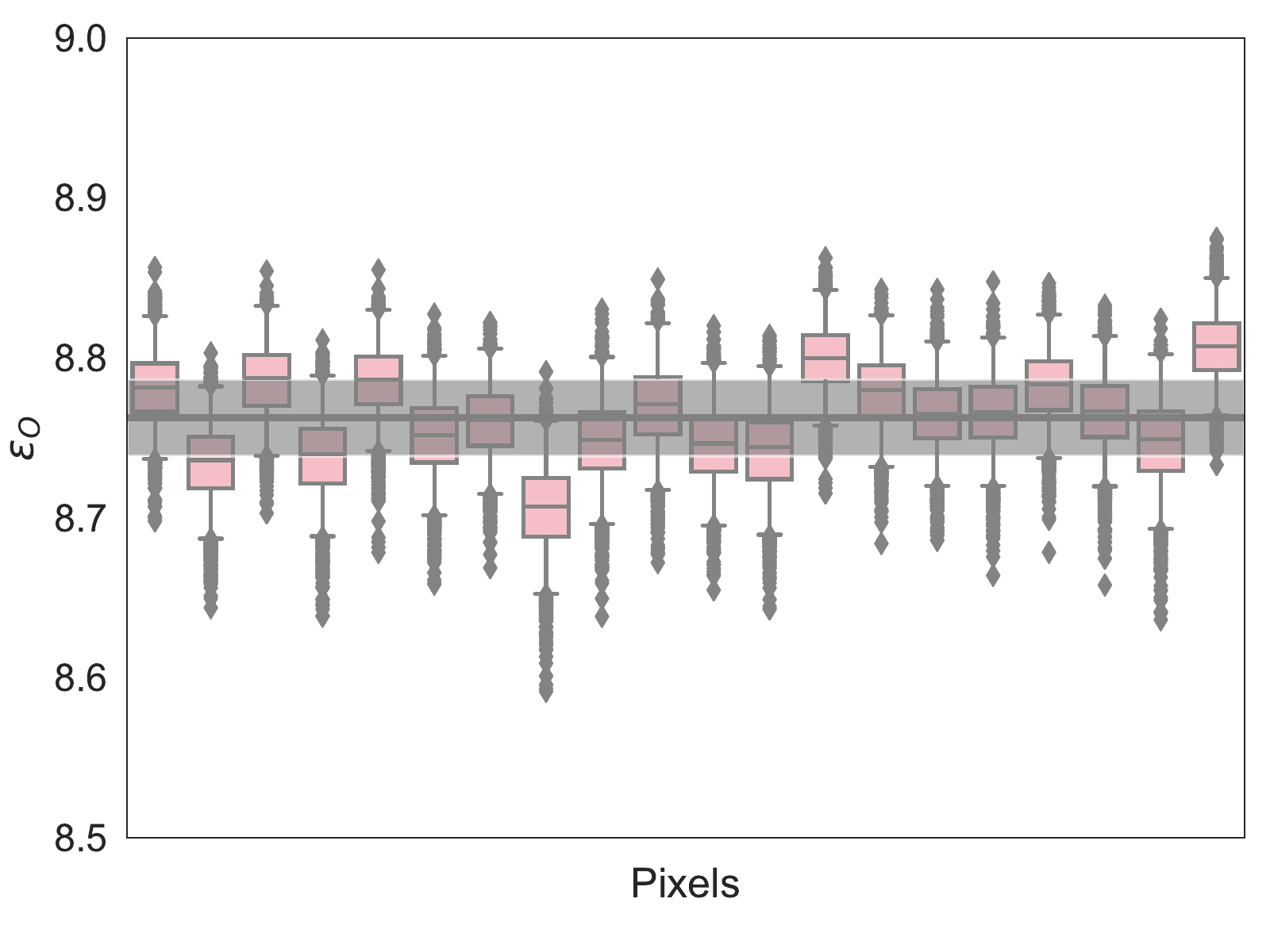}
\caption{Distribution of oxygen abundance for each pixel. Granules are shown in the \textit{upper panel} and lanes in the \textit{bottom panel}. The gray horizontal line represents the mean of the means of each pixel and the gray area represents the standard deviation. The horizontal line inside each color box represents the median value of the marginal posterior distribution, and the color box covers the range encompassing 50\% of the probability. }
\label{fig:ao_vs_pixel_mean}
\end{figure}

The marginal posterior distributions for the oxygen abundance are shown as boxplots in
Fig.~\ref{fig:ao_vs_pixel_mean}, where the granules are plotted in green (upper panel) 
and the lanes in pink (bottom panel). As usual in boxplots, the horizontal line inside 
the colored boxes 
represents the median value of the marginal posterior distributions, and the colored 
box covers from the first to the third quartile (amounting to 50\% probability). 
The lines going beyond the box extend to show the distribution up to two standard 
deviations. The remaining points are determined to be outliers. The horizontal 
gray line corresponds to the global mean (the mean of the individual means) and the gray area shows one standard deviation.

The results show that all 20 independently analyzed granules  yield values that are 
statistically consistent and compatible inside the error bars of each individual 
determination. This is also the case for the lanes. Moreover, granules and lanes show 
results that are compatible within approximately 2-$\sigma$. This is an indication of 
robustness, because there is no reason a priori to expect that all 40 spatial locations 
will give the same abundance. In that sense, this is a 
consistent determination of the photospheric abundance. The inferred abundance for 
oxygen in granules is $\log(\epsilon_O)=8.83\pm 0.02$, while it is 
$\log(\epsilon_O)=8.76 \pm 0.02$ for intergranular lanes. These values are obtained 
by computing the means of the individual means in each pixel. 

\begin{figure}
\centering
\includegraphics[trim={3 10 5 7},clip,width=.95\hsize]{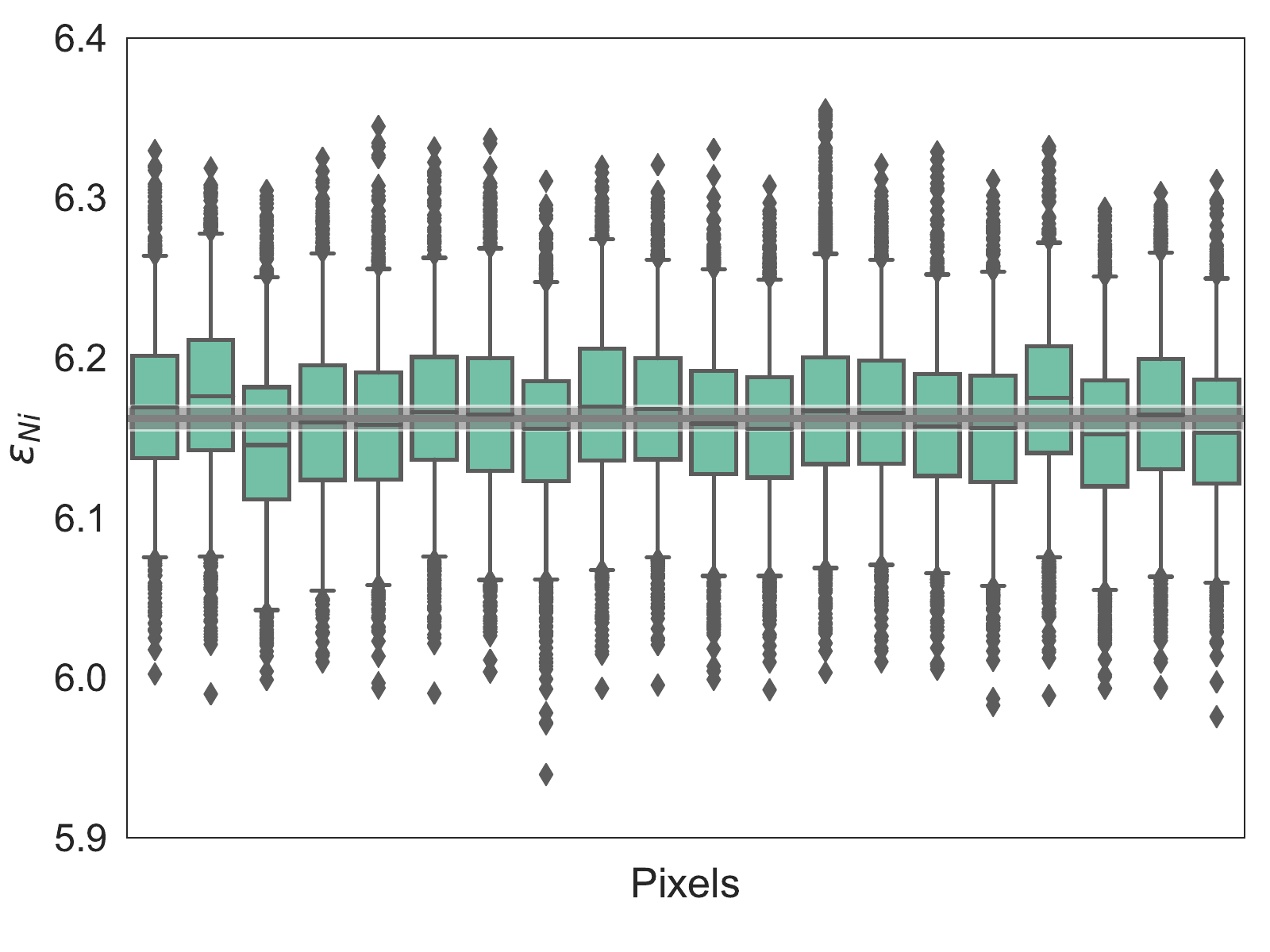}
\includegraphics[trim={3 10 5 7},clip,width=.95\hsize]{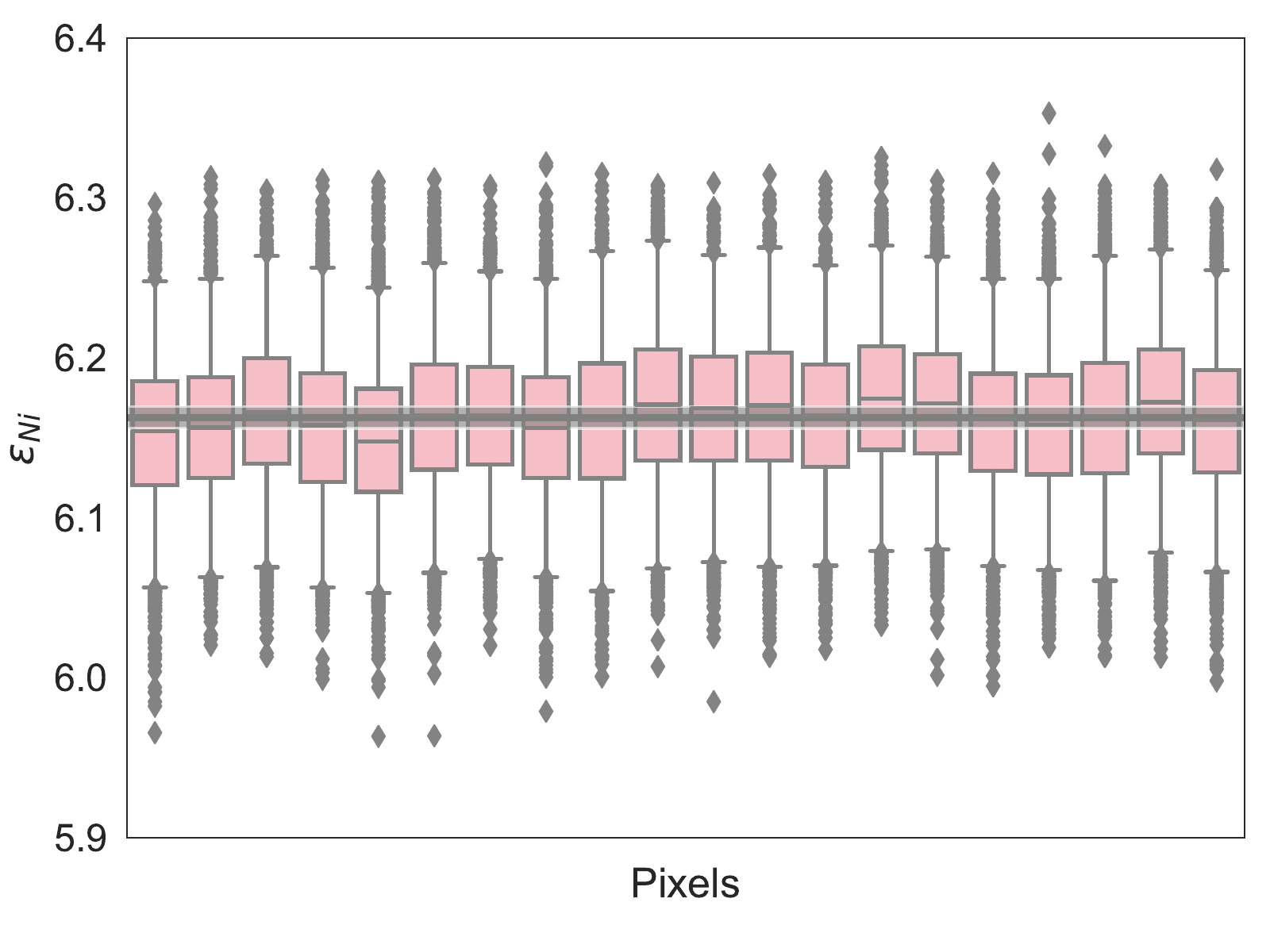}
\caption{Distribution of nickel abundances for each pixel. The gray horizontal line represents the mean of the means of each pixel and the gray area represents the standard deviation. Granules are shown in the \textit{upper panel} and lanes in the \textit{bottom panel}.}
\label{fig:ani_vs_pixel_mean}
\end{figure}

We interpret the small difference between the oxygen abundance obtained in granules and 
lanes as a consequence of some systematic errors in our study rather than a real abundance difference. For instance, errors in the determination of the temperature and velocity considerably affect
the line profile. If  the inversions are repeated, changing the number of 
nodes or some other parameters, the models obtained will be slightly different, resulting in a different oxygen abundance. Furthermore, there are also errors in the atomic parameters and other systematic effects beyond our control. All of these effects could explain the 2-$\sigma$ tension found between the granules and lanes.
 
For completeness, Fig.~\ref{fig:ani_vs_pixel_mean} displays the marginal posterior
distributions for the nickel abundance. Here, the marginal posterior distributions 
are very narrow because the data are extremely consistent with the priors and consistent
between granules and lanes. The mean 
of the means of the individual pixels is $\log(\epsilon_{Ni})=6.16 \pm 0.01$ both 
for the granules and for the lanes. 

\subsection{Hierarchical partial pooling model}

\begin{figure}
\centering
\includegraphics[trim={0 10 5 7},clip,width=.95\hsize]{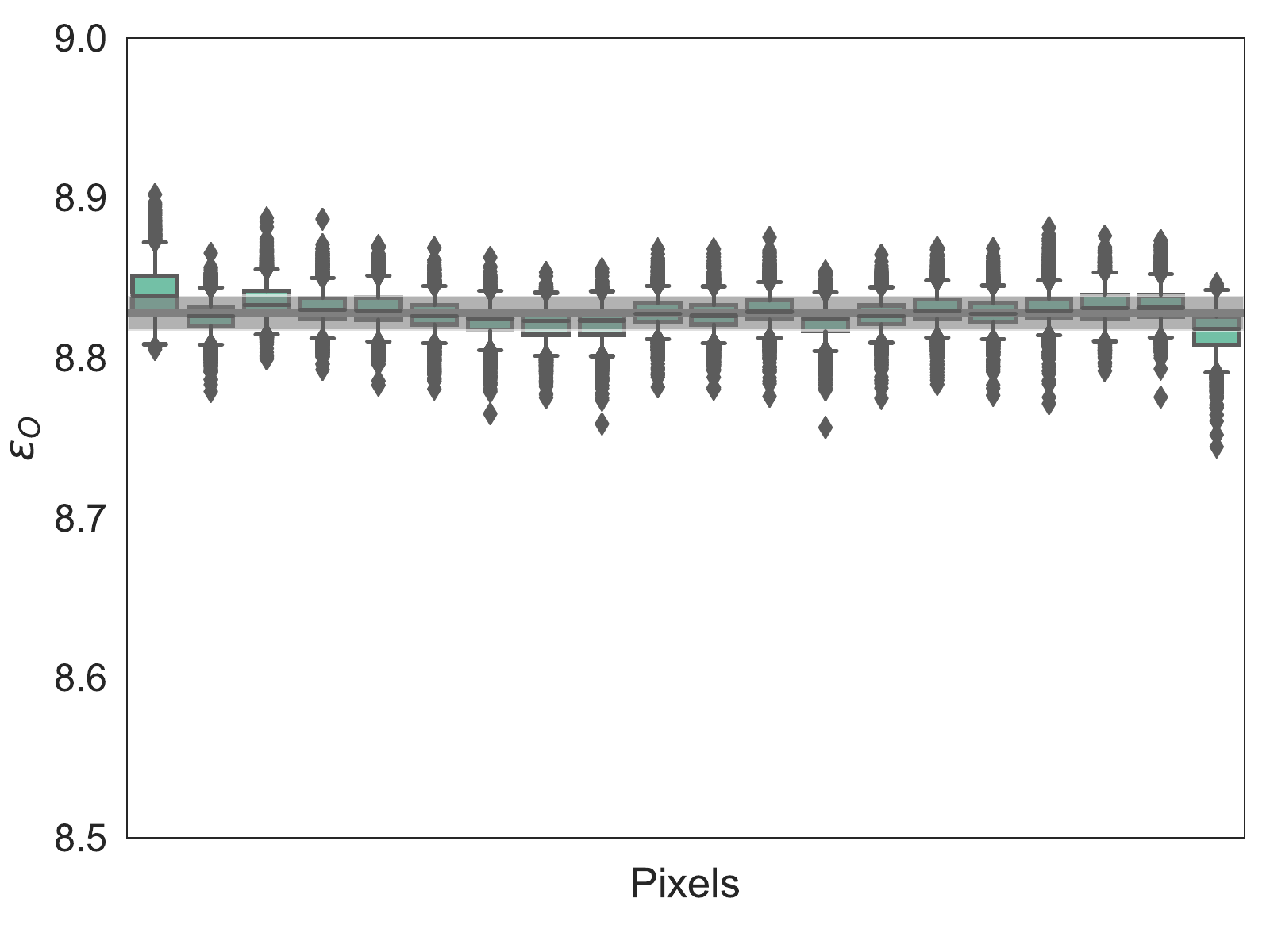}
\includegraphics[trim={0 10 5 7},clip,width=.95\hsize]{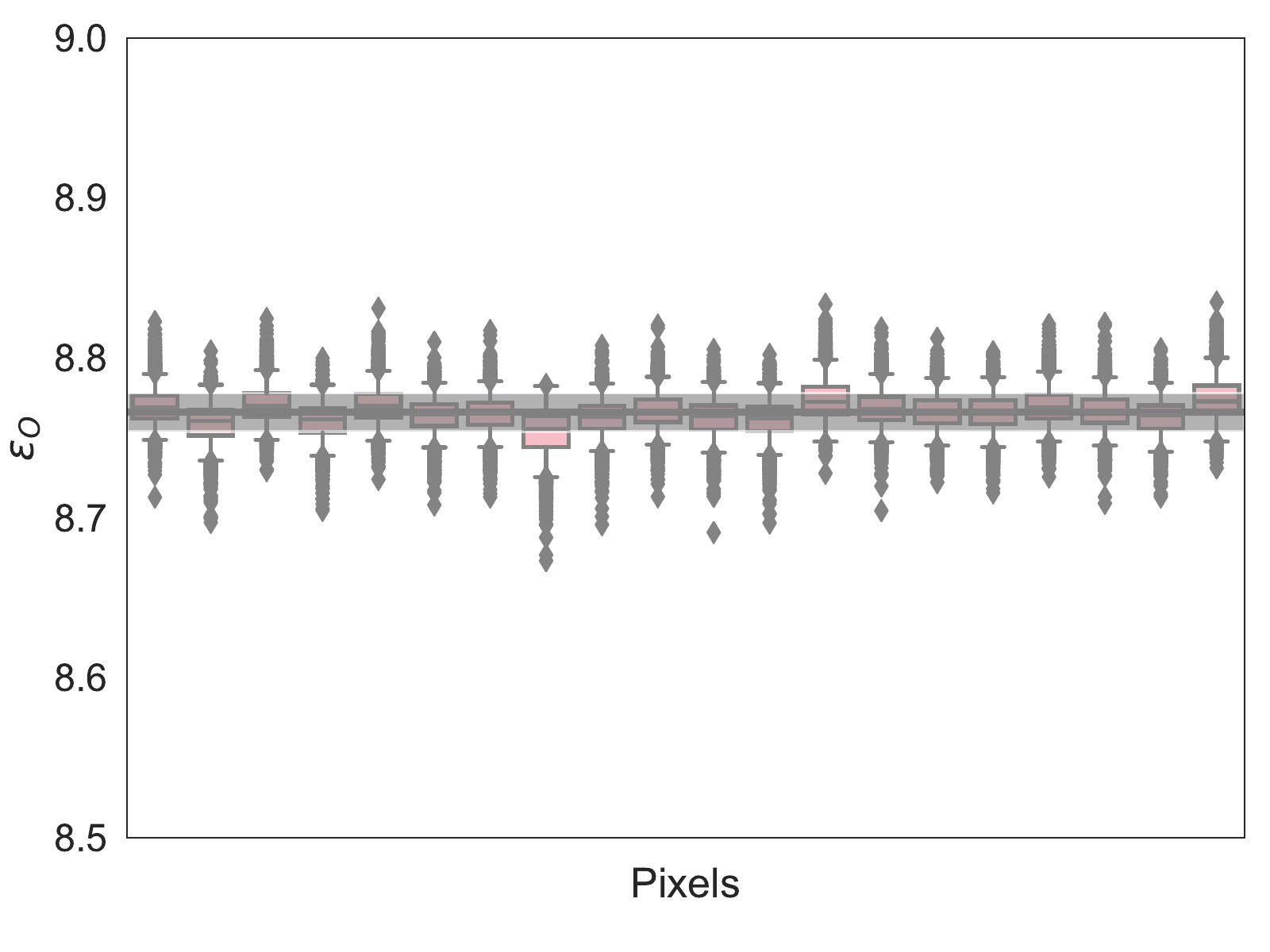}
\caption{Distribution of oxygen abundance for each pixel. The gray horizontal line represents the hyperparameter of oxygen abundance and the gray area covers the probable values within the standard deviation, which is the other hyperparameter. Granules are shown in the \textit{upper panel} and lanes in the \textit{bottom panel}.}
\label{fig:ao_vs_pixel_hier}
\end{figure}

\begin{figure*} 
\centering
\includegraphics[trim={5 7 10 8},clip,width=.9\hsize]{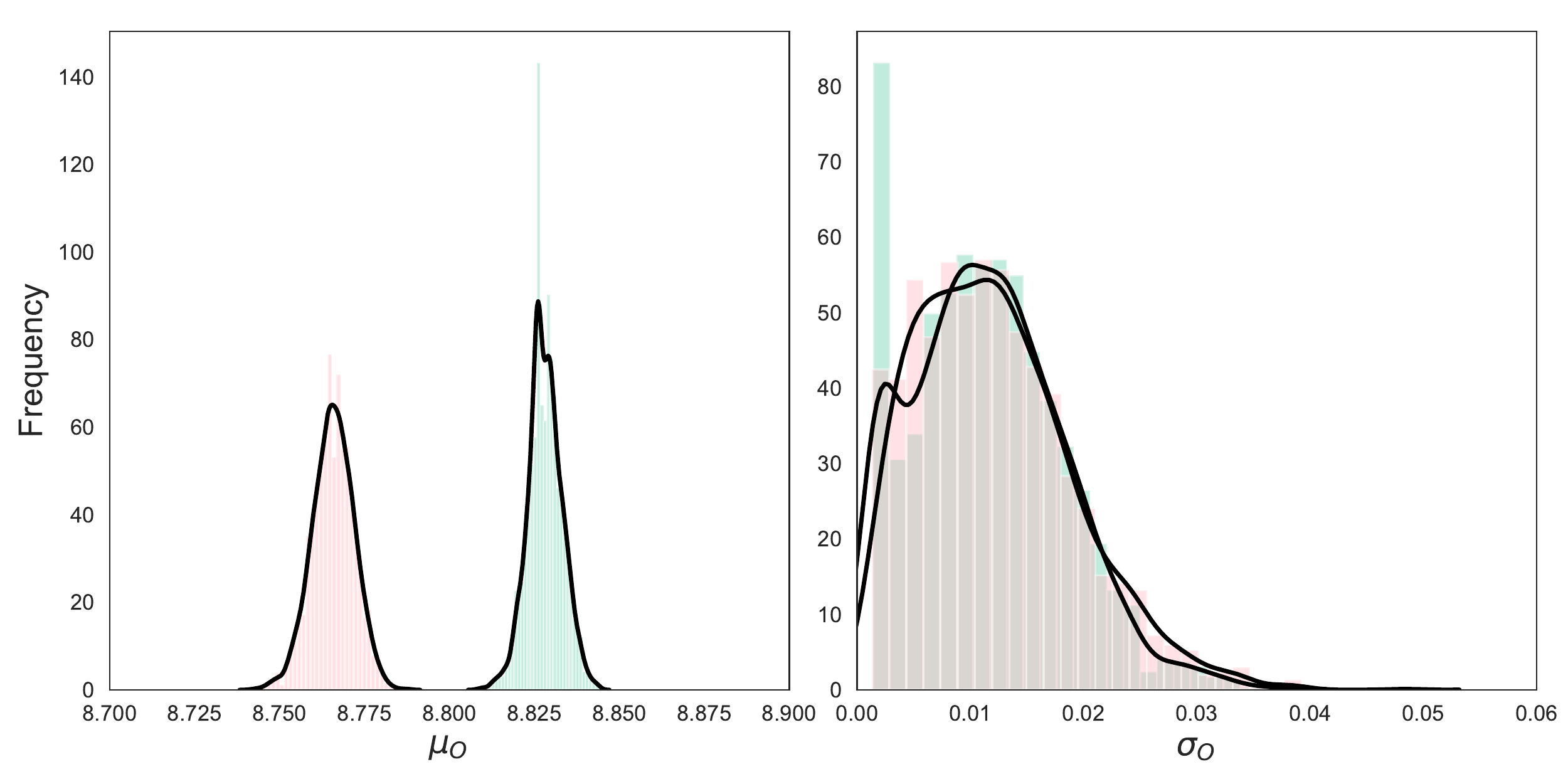}
\includegraphics[trim={5 7 10 8},clip,width=.9\hsize]{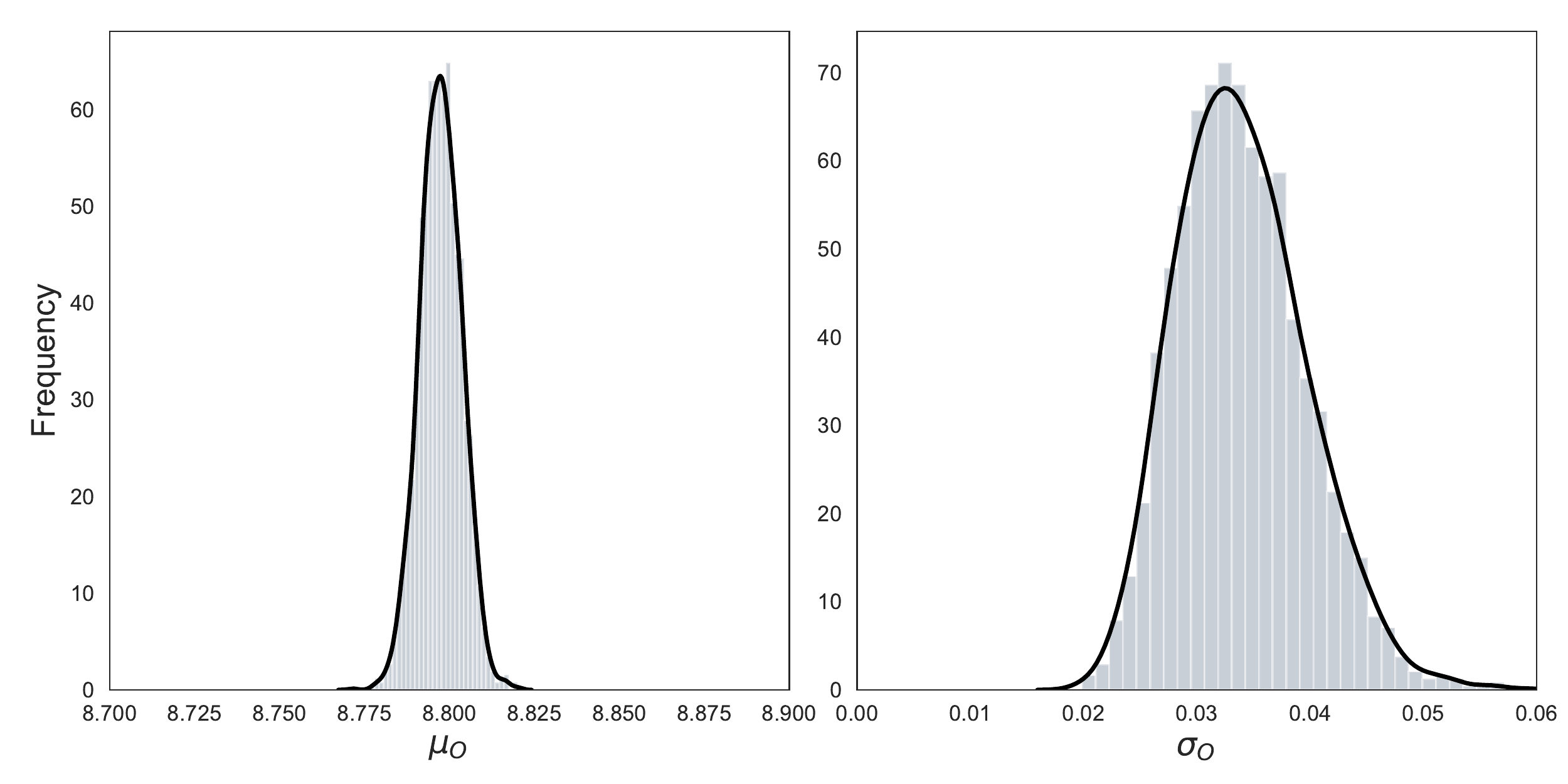}
\caption{Distributions of the hyperparameters for the hiearchical partial pooling model. The \textit{left panels} show the global oxygen abundance and the \textit{right panels} the standard deviation. Top panels show the granules (green) and lanes (pink) as treated separately and the bottom panels show the treatment of  all the pixels together.}
\label{fig:hier}
\end{figure*}

The results of the marginal posterior distribution for the oxygen abundance 
in the partial pooling model are shown in Fig.~\ref{fig:ao_vs_pixel_hier}. 
In this case, the horizontal line represents the mean of the marginal posterior for the
hyperparameter $\mu_O$, while the shaded area represents the range between $\mu_O \pm \sigma_O$. 
The hierarchical model displays a strong shrinkage effect as compared with the previous
model, something that is characteristic of hierarchical Bayesian models. The estimated
values for the oxygen abundance for all pixels are pulled towards the group mean, 
with a much smaller dispersion. In this case, we find $\mu_O$=8.83$\pm$0.01 and $\mu_O$=8.77$\pm$0.01 for the
granules and the lanes, respectively. On the other hand, the $\sigma_O$
distribution has a mean of 0.01 for both granules and lanes; where  95\% of the distribution has values of up to $\sigma_O$<0.023 for the granules and
$\sigma_O$<0.025 for the lanes (see Fig.~\ref{fig:hier}).

\begin{figure} 
\centering
\includegraphics[trim={0 10 5 7},clip,width=.95\hsize]{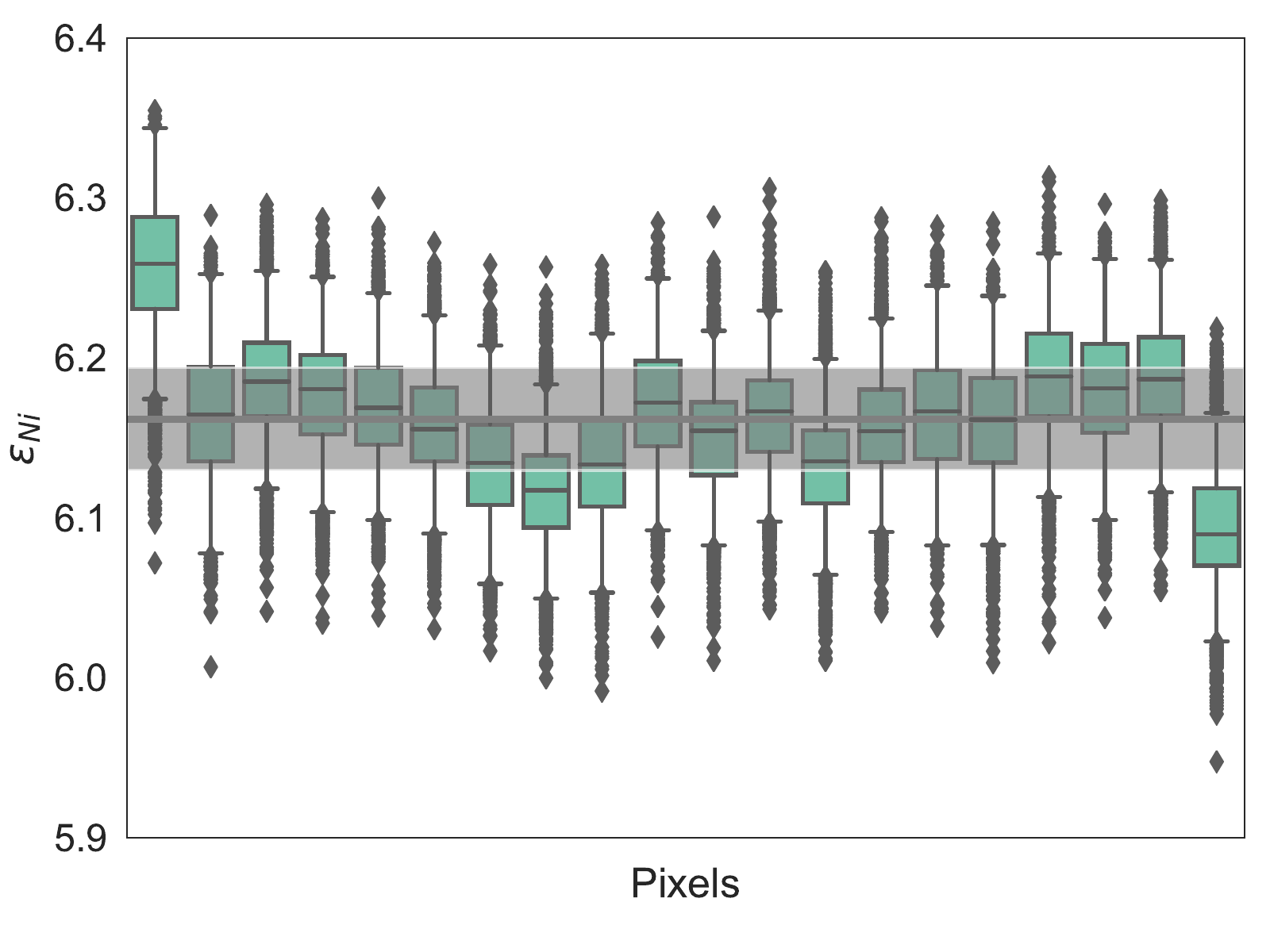}
\includegraphics[trim={0 10 5 7},clip,width=.95\hsize]{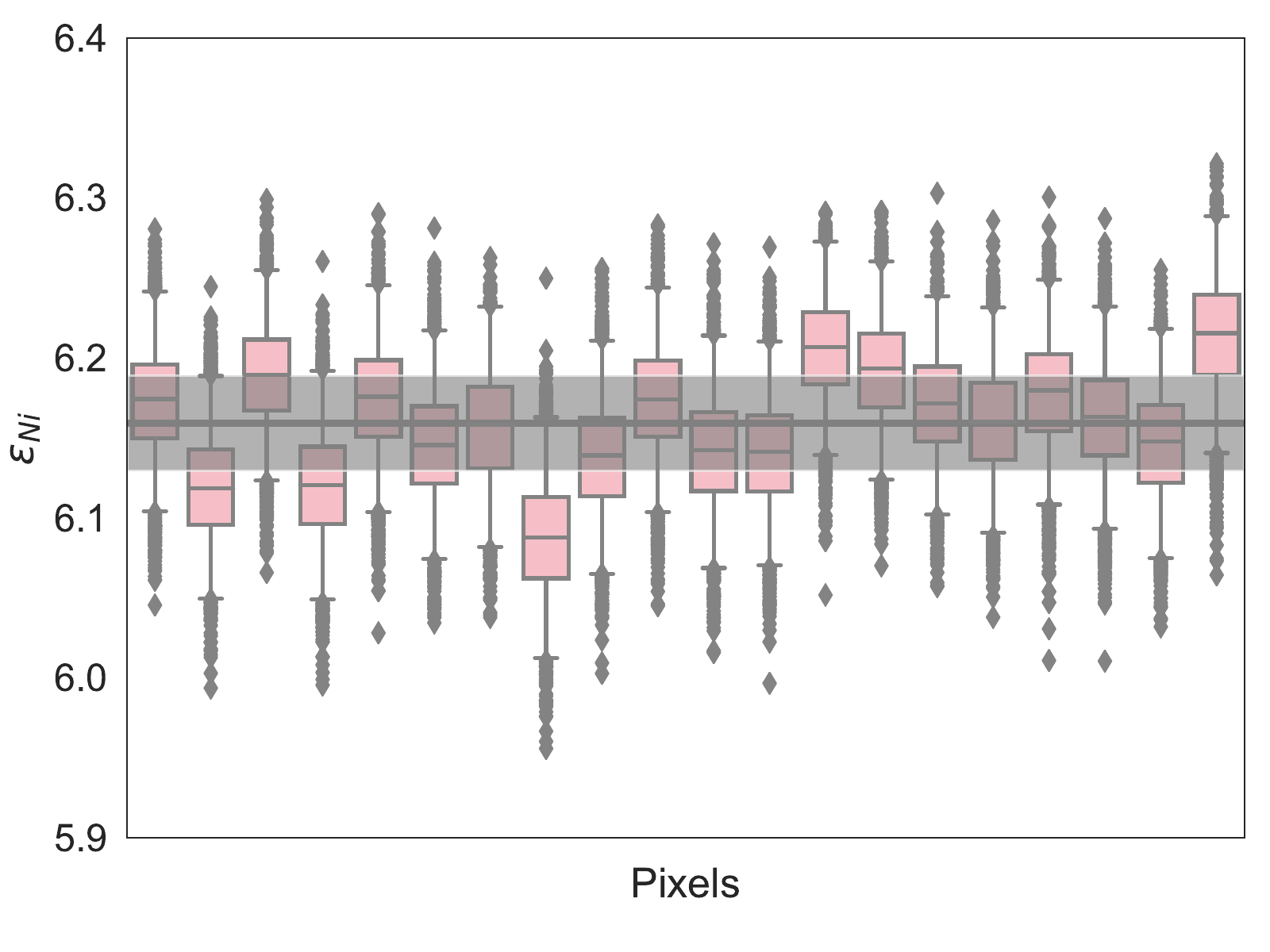}
\caption{Distribution of nickel abundance for each pixel. The gray horizontal line represents the mean value of the mean values for the nickel abundance in each pixel and the gray area covers the probable values within the standard deviation. Granules are shown in the \textit{upper panel} and lanes in the \textit{bottom panel}.}
\label{fig:ani_vs_pixel_hier}
\end{figure}

The marginal posterior distributions for the nickel abundance in this case are shown in
Fig.~\ref{fig:ani_vs_pixel_hier}. The mean value obtained is $\log(\epsilon_{Ni})=6.16 \pm 0.03$,
both for the granules and for the lanes. In general, we find similar values in both the
unpooled and partially pooled model for the oxygen and nickel abundances, which 
demonstrates the robustness of our result.

\begin{figure} 
\centering
\includegraphics[trim={0 10 5 7},clip,width=.99\hsize]{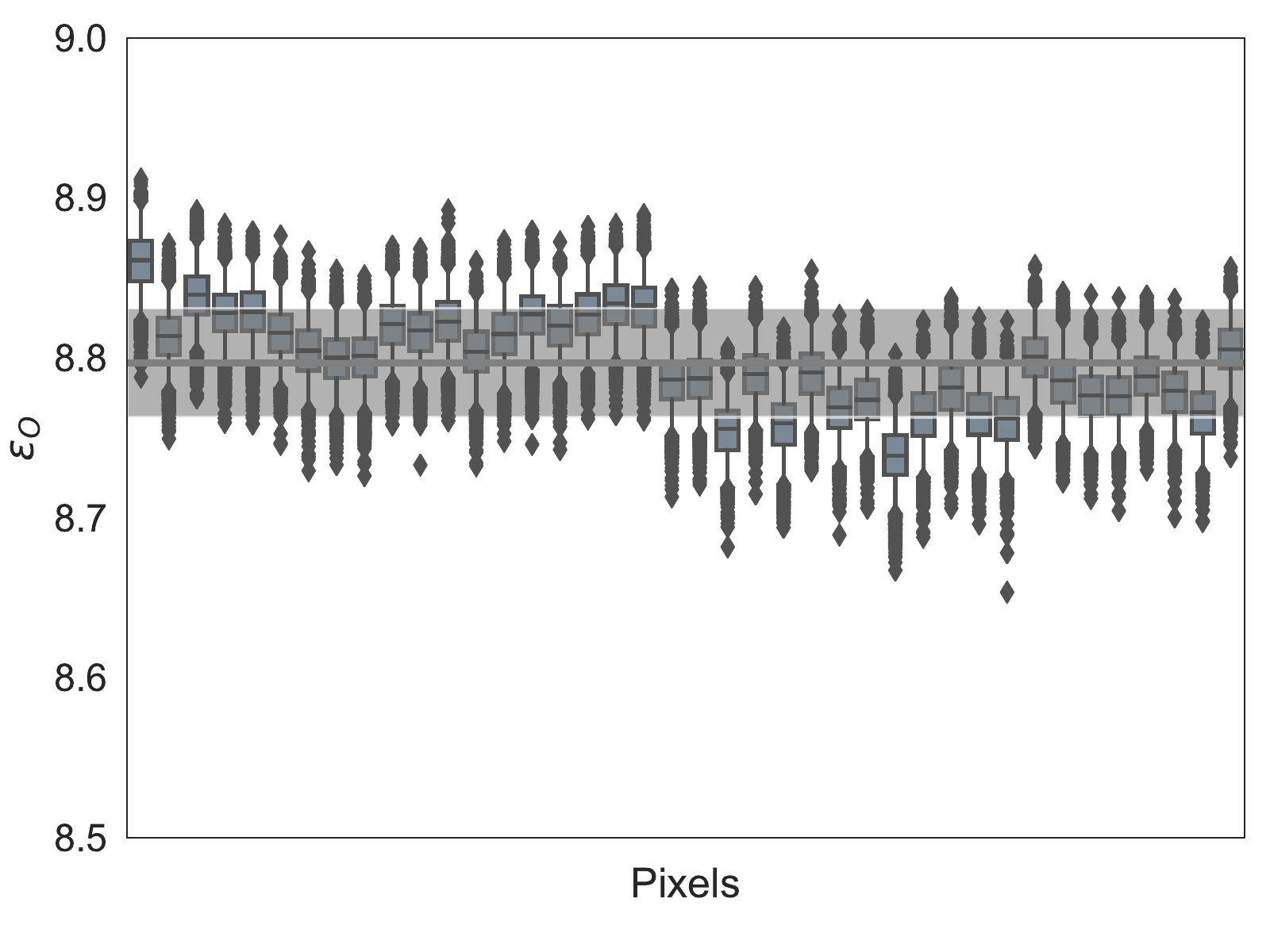}
\includegraphics[trim={0 10 5 7},clip,width=.99\hsize]{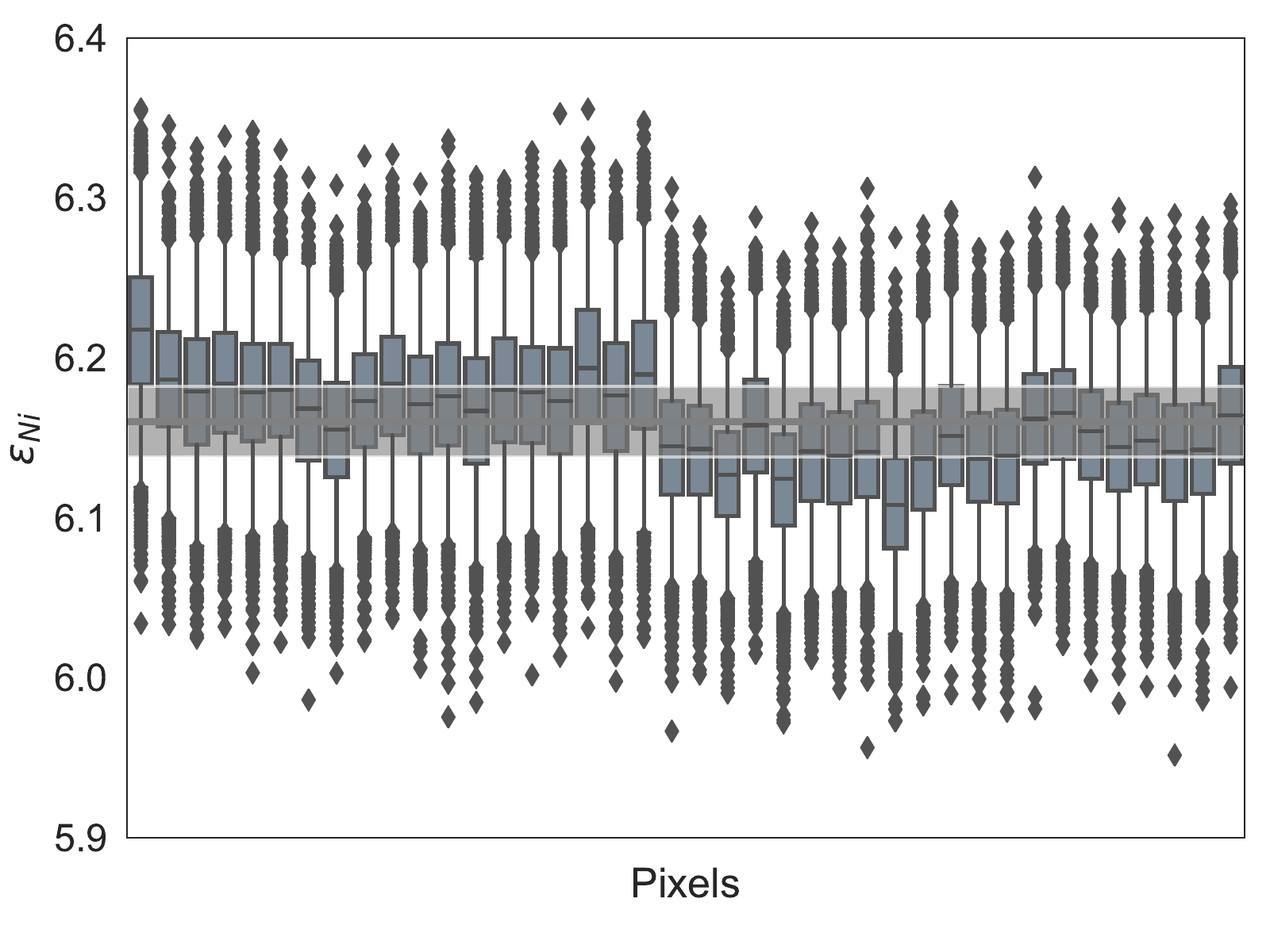}
\caption{Distribution of oxygen abundance (\textit{upper panel}) and nickel abundance (\textit{bottom panel}) for each pixel considering granules and lanes together. For the oxygen abundance, the gray horizontal line represents the hyperparameter of oxygen abundance and the gray area covers the probable values within the standard deviation, which is the other hyperparameter. For the nickel abundance, the gray horizontal line is the mean of the means of nickel abundance in each pixel and the gray area covers the probable values within the standard deviation.}
\label{fig:hier_all}
\end{figure}

A final experiment consists of using a hierarchical partial pooling model with granules
and lanes together. The aim is to test whether or not the differences in the results 
found when granules and lanes are treated separately are model dependent. For this purpose, we use the same 
prior distribution both for granules and lanes. Figure~\ref{fig:hier_all} shows the marginal
posteriors. The value inferred for the nickel abundance is still the same as that
of previous models, namely $\log(\epsilon_{Ni})=6.16 \pm 0.02$, with a slightly smaller uncertainty. 
For the oxygen abundance we obtain a distribution where the mean value is between 
the values that we obtained for granules and lanes, namely $\mu_O=8.80 \pm 0.01$ and 
$\sigma_O =0.03 \pm 0.01$.

Finally, Fig.~\ref{fig:hier} shows the marginal posterior distributions 
for the hyperparameters of the oxygen abundance prior. The upper panels correspond
to the hierarchical model of granules and lanes separately while the lower
panels refer to the case in which granules and lanes are considered together.
It is clear that the marginal posterior for $\mu_O$ for the lower panel lies
somewhere between those of the upper panel. Concerning $\sigma_O$, it is clear
that marginal posterior peaks at larger values when both granules and lanes are
considered together. However, the width of the posterior seems to be similar in
the two inferences. 

In summary, our currently recommended value for the oxygen abundance
based on the results of this work is that from the hierarchical model in which both granules and lanes
are considered together, which gives $\log(\epsilon_O)=8.80 \pm 0.03$. This value is compatible with the results obtained by \cite{1998SSRv...85..161G} and the results that we achieved in a previous work \citep[see][]{2017A&A...600A..45C}. Therefore, we infer an oxygen abundance comparable with the old high-Z abundances \citep[see e.g.,][]{1998SSRv...85..161G}.

\section{Summary and conclusions}

The solar oxygen abundance remains an unresolved and problematic issue
in astrophysics, not only because it is a crucial element (very
important for the construction of solar interior models and because
other elements are measured relative to it) but also because it is
extremely difficult to measure with sufficient accuracy. \cite{2015A&A...577A..25S} found that, with very minor tweaks in the data processing, it
is possible to obtain (from the same observations) a whole range of
values as broad as $\log(\epsilon_O) \in [8.7, 8.9]$, but this latter author did not explore the
relative likelihood of all the possible outcomes. Random errors
arising from uncertainties in the observations, the atomic parameters,
radiative transfer, and calibration (especially wavelength grid and
continuum level) are relatively straightforward to estimate. A full
Bayesian treatment of these errors was provided in \cite{2017A&A...600A..45C}.

However, the most important challenge is in finding a way to estimate the
systematic errors, particularly those introduced by the (prescribed)
atmosphere. Thus far, all previous determinations start from a given
model atmosphere and then proceed by adjusting the abundances until an
optimal fit to the observations is attained. The model is considered
perfect and it is impossible to asses to what extent the uncertainties
in the model affect the abundances obtained.

This work presents a novel approach, taking advantage of the fact that
the solar surface is spatially resolved. We perform 40 independent
inferences at different solar locations with varying thermal, dynamic,
and magnetic conditions. Ideally, all 40 abundances should converge to
the same value. However, imperfections in the modeling will produce a
spread in the results from which both random and systematic
uncertainties may be estimated. We observe such a spread at a level of
0.03~dex.

The Bayesian analysis with a hierarchical model assimilates all the
available data to produce a final result of $\log(\epsilon_O)=8.80 \pm 0.03$,
which is consistent with the results of granules and lanes taken
separately (8.83$\pm$0.02 and 8.76$\pm$0.02, respectively). The fact
that the most probable values for granules and lanes are slightly
different (at the level of 2-$\sigma$) is an indication that some
systematic errors may still linger in the analysis. Such systematic error
might arise from the data calibration, NLTE effects in the FeI lines
employed in the inversions, model uncertainties, or even from errors in
the atomic parameters.

\begin{acknowledgements}
We acknowledge financial support from the Spanish Ministerio de Ciencia, Innovaci\'on y 
Universidades through project PGC2018-102108-B-I00 and FEDER funds.
M.C.A. acknowledges Fundaci\'on  la  Caixa  for  the  financial  support  received  in  the  
form  of a PhD contract. We thank the anonymous referee for helpful comments and suggestions
that helped improve an earlier version of the manuscript. This research has made use of NASA's Astrophysics Data System 
Bibliographic Services. This work has made use of the VALD database, operated at Uppsala 
University, the Institute of Astronomy RAS in Moscow, and the University of Vienna. The 
Vacuum Tower Telescope in Tenerife is operated by the Kiepenheuer-Institut f\"ur Sonnenphysik (Germany) 
in the Spanish Observatorio de Iza\~na of the Instituto de Astrof\'isica de Canarias. We acknowledge 
the community effort devoted to the development of the open-source Python packages that were used in this work. 
      
\end{acknowledgements}
%
%
%

\begin{appendix} 
\section{Posterior predictive checks for the partial pooling models}

\begin{figure*} 
\centering
\includegraphics[trim={0cm 0cm 0cm 0cm},clip,width=0.86\hsize]{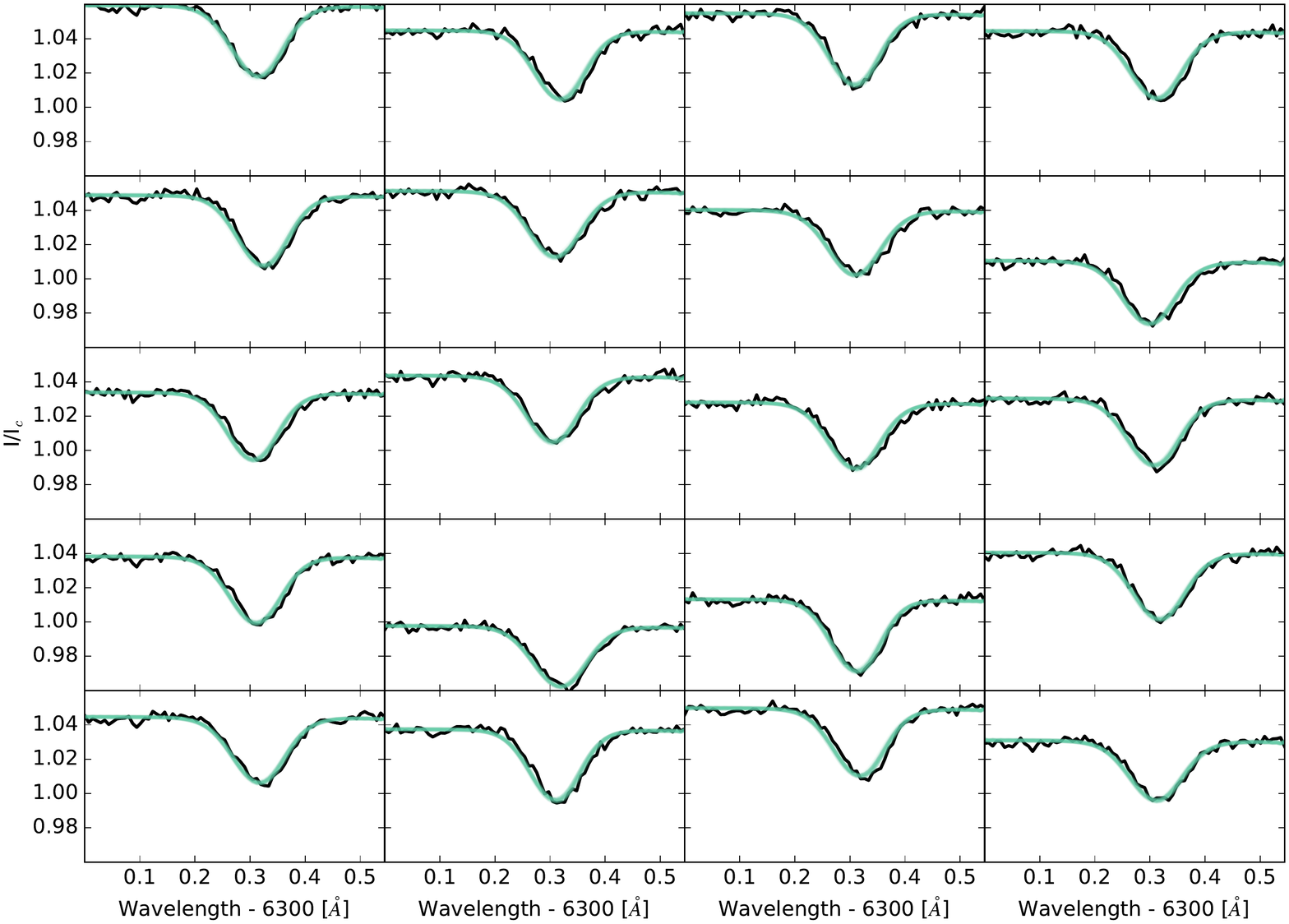} 
\caption{Same as Fig.~\ref{fig:bay_gran} but for the hierarchical partial pooling model.}
\label{fig:bay_gran_hier}
\end{figure*}

\begin{figure*} 
\centering
\includegraphics[trim={0cm 0cm 0cm 0cm},clip,width=0.86\hsize]{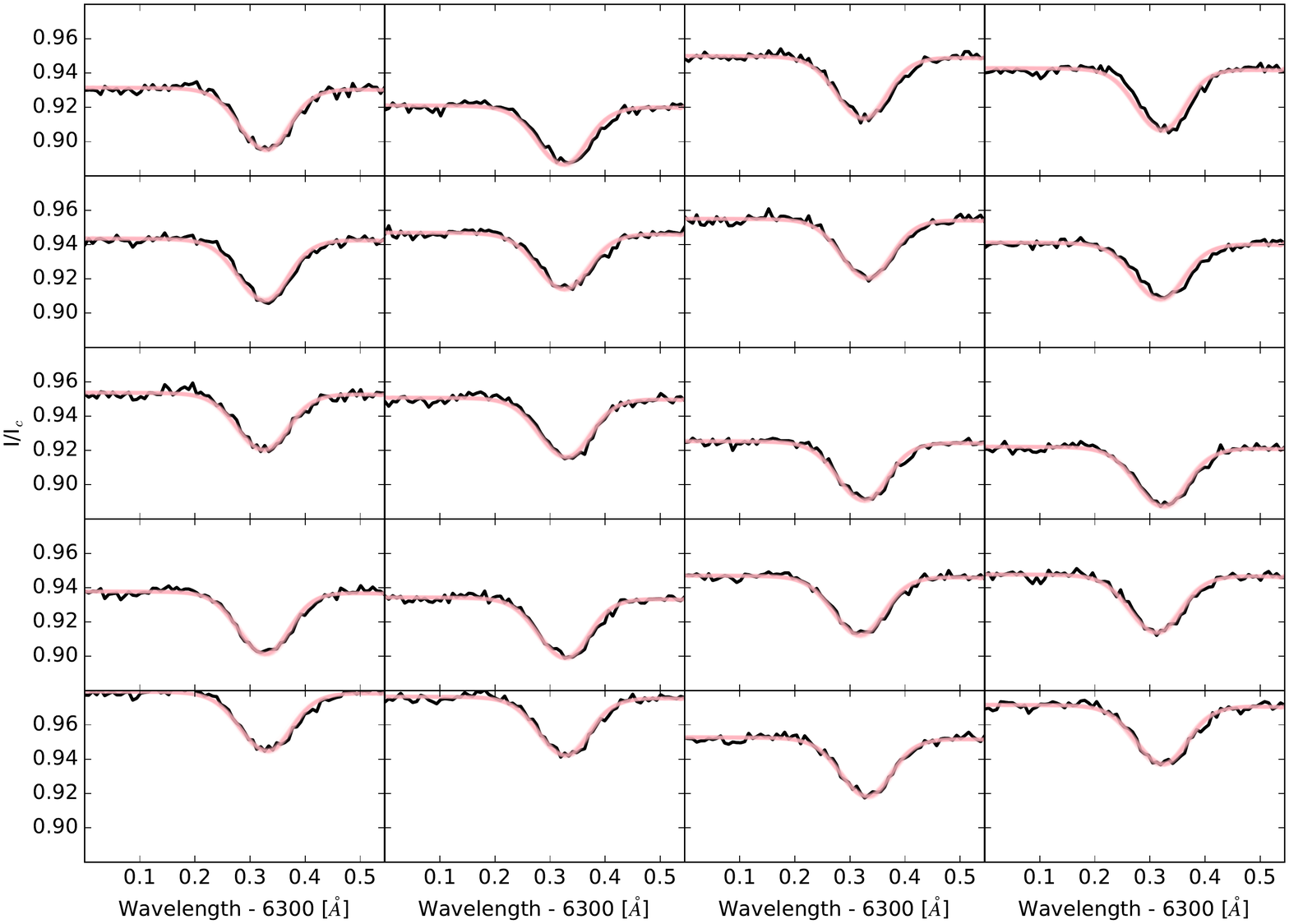} 
\caption{Same as Fig.~\ref{fig:bay_lane} but for the hierarchical partial pooling model.}
\label{fig:bay_lane_hier}
\end{figure*}

\begin{figure*} 
\centering
\includegraphics[trim={0cm 0cm 0cm 0cm},clip,width=0.86\hsize]{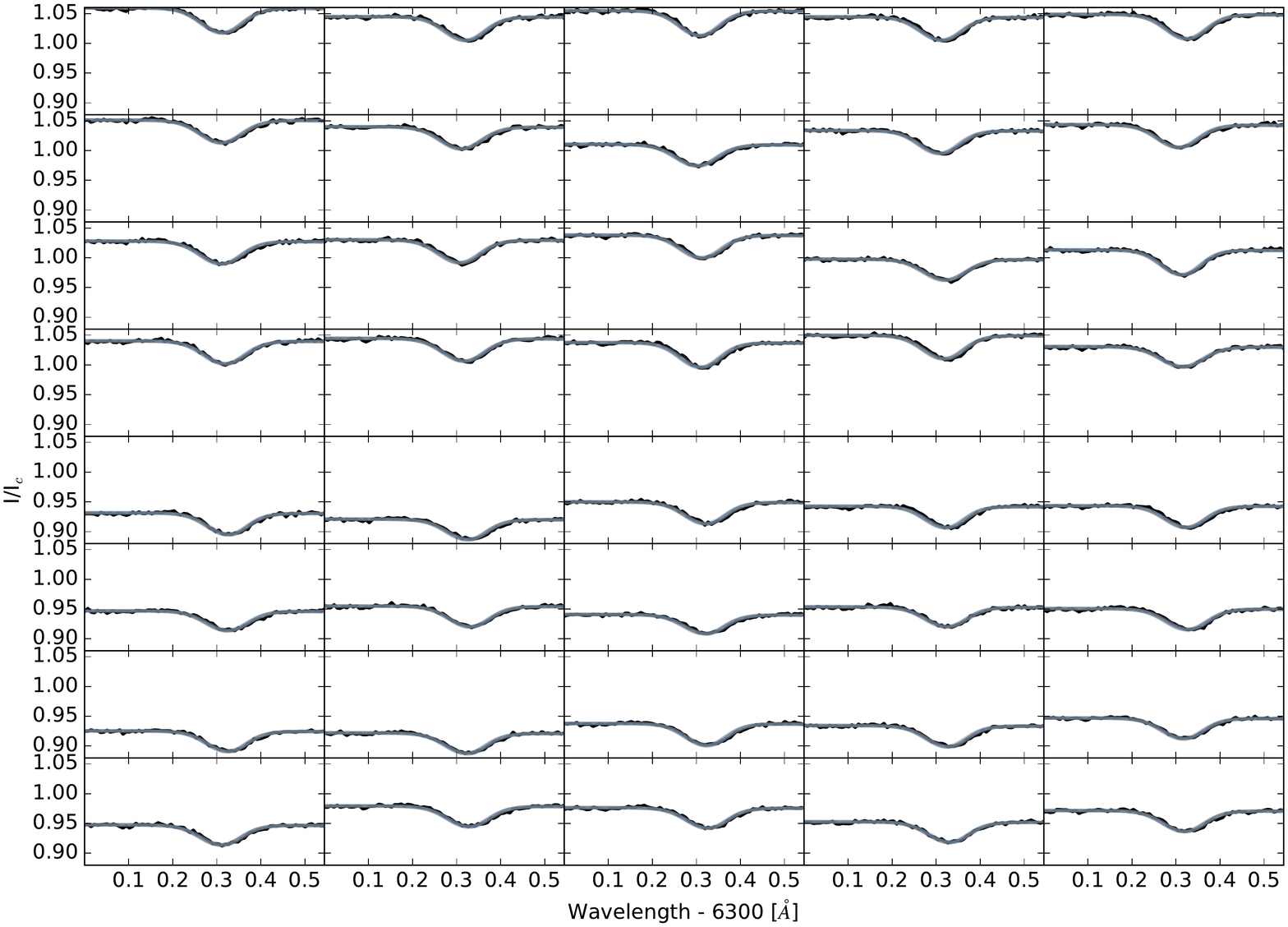} 
\caption{Same as Fig.~\ref{fig:bay_gran_hier} and Fig.~\ref{fig:bay_lane_hier} but considering the granules and the lanes together in the Bayesian inference.}
\label{fig:bay_hier_all}
\end{figure*}

Here we show the posterior predictive checks, in which the synthetic [O~{\sc i}] line is obtained from different samples from the posterior in the hierarchical partial pooling models; Figure~\ref{fig:bay_gran_hier} for the granules, and Fig.~\ref{fig:bay_lane_hier}  for the lanes. Finally, Fig.~\ref{fig:bay_hier_all} is for the hierarchical partial pooling model considering the granules and the lanes together. 


\end{appendix}

\end{document}